\documentclass[aps,prd,superscriptaddress,nofootinbib,preprint]{revtex4}
\pdfoutput=1
\usepackage[utf8]{inputenc}
\usepackage[english]{babel}
\usepackage{amsmath}
\usepackage{subfigure}
\usepackage{amsfonts}
\usepackage{amssymb}
\usepackage{graphicx}
\usepackage{color}
\newcommand{\be}{\begin{equation}}
\newcommand{\ee}{\end{equation}}
\newcommand{\bea}{\begin{eqnarray}}
\newcommand{\eea}{\end{eqnarray}}

\newcommand{\umt}{{\rm U(1)}_{L_{\mu}-L_{\tau}}}
\newcommand{\vmt}{v_{\mu \tau}}
\newcommand{\gmt}{g_{\mu \tau}}
\newcommand{\zmt}{Z_{\mu \tau}}
\newcommand{\zbl}{{Z_{\mu\tau}}}
\newcommand{\mzmt}{M_{Z_{\mu\tau}}}

\newcommand{\smgauge}{{\rm SU}(2)_{L}\times {\rm U}(1)_{Y}}
\newcommand{\nn}{\nonumber}

\catcode`@=12
\def\la{\mathrel{\mathchoice {\vcenter{\offinterlineskip\halign{\hfil
$\displaystyle##$\hfil\cr<\cr\sim\cr}}}
{\vcenter{\offinterlineskip\halign{\hfil$\textstyle##$\hfil\cr<\cr\sim\cr}}}
{\vcenter{\offinterlineskip\halign{\hfil$\scriptstyle##$\hfil\cr<\cr\sim\cr}}}
{\vcenter{\offinterlineskip\halign{\hfil$\scriptscriptstyle##$\hfil\cr<\cr\sim
\cr}}}}}

\begin{document}
\title{Explaining the 3.5 keV X-ray Line in a ${L_{\mu}-L_{\tau}}$ Extension of the Inert Doublet Model}

\author{Anirban Biswas\footnote{Presently at Department of Physics,
IIT Guwahati, Guwahati, Assam, India 781039}}
\email{anirban.biswas.sinp@gmail.com}
\affiliation{Harish-Chandra Research Institute, HBNI, Chhatnag Road,
Jhunsi, Allahabad 211 019, India}
\author{Sandhya Choubey}
\email{sandhya@hri.res.in}
\affiliation{Harish-Chandra Research Institute, HBNI, Chhatnag Road,
Jhunsi, Allahabad 211 019, India}
\affiliation{Department of Physics, School of
Engineering Sciences, KTH Royal Institute of Technology, AlbaNova
University Center, 106 91 Stockholm, Sweden}
\author{Laura Covi}
\email{laura.covi@theorie.physik.uni-goettingen.de}
\affiliation{Institute for Theoretical Physics, Georg-August
University Göttingen, Friedrich-Hund-Platz
1, Göttingen, D-37077 Germany}
\author{Sarif Khan}
\email{sarifkhan@hri.res.in}
\affiliation{Harish-Chandra Research Institute, HBNI, Chhatnag Road,
Jhunsi, Allahabad 211 019, India}

\begin{abstract}
We explain the existence of neutrino masses and their flavour structure, dark matter 
relic abundance and the observed 3.5 keV X-ray line within the 
framework of a gauged $U(1)_{L_{\mu} - L_{\tau}}$ extension of the 
``scotogenic" model. 
In the $U(1)_{L_{\mu} - L_{\tau}}$ 
symmetric limit, two of the RH neutrinos are degenerate in mass, while the 
third is heavier. 
The $U(1)_{L_{\mu} - L_{\tau}}$ symmetry is broken 
spontaneously. 
Firstly, this breaks the $\mu-\tau$ symmetry in the light neutrino sector. 
Secondly, this results in mild splitting of the two degenerate RH neutrinos, with their mass 
difference given in terms of the $U(1)_{L_{\mu} - L_{\tau}}$ breaking parameter. 
Finally, we get a massive $Z_{\mu\tau}$ gauge boson. 
Due to the added $Z_2$ symmetry under which the RH neutrinos and the inert 
doublet are odd, the canonical Type-I seesaw is forbidden and the tiny neutrino
masses are generated radiatively at one loop. The same $Z_2$ symmetry also ensures 
that the lightest RH neutrino is stable and the other two can only decay into the lightest one.
This makes the two nearly-degenerate lighter neutrinos a two-component dark matter, which 
in our model are produced by the freeze-in mechanism via the decay of the $Z_{\mu\tau}$ 
gauge boson in the early universe. 
We show that the next-to-lightest RH neutrino has a very long 
lifetime and decays into the lightest one at the present epoch 
explaining the observed 3.5 keV line. 
\end{abstract}

\maketitle
\newpage

\section{Introduction}

Experimental proof of non-zero neutrino masses and mixing as well as the dark matter in the universe remain the two most compelling evidences of the existence of physics beyond the standard model. Different neutrino oscillation experiments have confirmed the existence of flavour oscillations which can be explained only if neutrinos have tiny masses and mixing \cite{Fukuda:1998mi, Ahmad:2002jz, Eguchi:2002dm}. 
On the other hand, observations of the flatness of the galaxy rotation curve \cite{Sofue:2000jx}, gravitational lensing \cite{Bartelmann:1999yn}, cosmic microwave background anisotropy \cite{Hinshaw:2012aka,Ade:2015xua} and more recently the observation of bullet cluster by NASA's Chandra Satellite \cite{Harvey:2015hha} demand that there must a non-baryonic component of matter in the universe, usually referred to as dark matter (DM).  One of the most promising particle DM candidate is the Weakly Interacting Massive Particles (WIMP). However, the null result from the different earth and satellite-based direct and indirect DM searches have put severe constraints on the WIMP paradigm \cite{Akerib:2017kat, Aprile:2017iyp}. One of the popular ways of shaking off the constraints from the direct and indirect DM searches is to postulate that the interaction strength of the DM with the standard model particles is extremely feeble. Such DM candidates go by the generic name Feebly Interacting Massive Particles (FIMP) \cite{Hall:2009bx, Yaguna:2011qn, Arcadi:2013aba, Molinaro:2014lfa,Biswas:2015sva, Merle:2015oja, Shakya:2015xnx, Biswas:2016bfo,Konig:2016dzg,Biswas:2016yjr, Biswas:2016iyh, Biswas:2017tce, Bernal:2017kxu} . Since their coupling with the standard model particles is feeble, they remain out of thermal equilibrium during the early universe when they are produced. Hence, these non-thermal particles  are produced by the so-called freeze-in \cite{Hall:2009bx} mechanism instead of the freeze-out process which produces thermal relics. 

More recently, the observation of an unknown 3.5 keV X-ray line 
in galaxies clusters~\cite{Bulbul:2014sua,Boyarsky:2014jta} 
and from the Galactic centre (GC) has been under much debate \cite{Cappelluti:2017ywp}. 
This excess has been confirmed by both the Chandra as well as NuSTAR satellites~\cite{Cappelluti:2017ywp}. 
It has been argued that this signal can come from iron line background and S XVI charge exchange.
Also such line has not been observed instead in stacked dwarf spheroidal
galaxies~\cite{Malyshev:2014xqa}, nor in galaxy spectra~ \cite{Anderson:2014tza}.
Nevertheless this signal has excited a lot of theoretical activity and can be
explained by a plethora of theoretical models~\cite{Chakraborty:2014tma,
Ishida:2014dlp, Finkbeiner:2014sja, 
Higaki:2014zua, Jaeckel:2014qea, Lee:2014xua, Kong:2014gea, Choi:2014tva, Baek:2014qwa, 
Tsuyuki:2014aia, Bezrukov:2014nza, Kolda:2014ppa, Allahverdi:2014dqa, Babu:2014pxa, 
Dudas:2014ixa, Aisati:2014nda, Modak:2014vva, Queiroz:2014yna,
Chiang:2014xra, Faisel:2014gda,
Falkowski:2014sma, Patra:2014sua, Arcadi:2014dca, DuttaBanik:2016jzv,
Abazajian:2017tcc, Heeck:2017xbu, Roszkowski:2017nbc, Bae:2017dpt,
Cosme:2017cxk, Brdar:2017wgy}.
Generically it points towards a very weakly interacting Dark Matter like a light sterile
neutrino, decaying into active neutrino and photon~\cite{Pal:1981rm}, but it can also 
arise from heavier DM particles in presence of mass degeneracy or from DM annihilation.


We address the issue of the observed neutrino masses and mixing, dark matter abundance of the universe and 
the 3.5 keV line within a BSM (Beyond SM) model, where we
have naturally a two component Dark Matter and a nearly degenerate long-lived state. 
We extend the SM gauge group by an anomaly free local $U(1)_{L_{\mu} - L_{\tau}}$ symmetry~\cite{He:1990pn, He:1991qd, Ma:2001md, Choubey:2004hn}. We break this gauge symmetry spontaneously by introducing in the model a SM singlet scalar charged under $U(1)_{L_{\mu} - L_{\tau}}$. The mass of the resultant neutral gauge boson is given in terms of the new gauge coupling and vacuum expectation value (VEV) of this scalar. Also included in the model are three RH neutrinos and a SM (inert) doublet scalar, both of which carry $-1$ charge with respect to an additional $Z_2$ symmetry, while all other particles carry charge $+1$. This forbids all Yukawa couplings of this doublet with the SM fermions (thereby earning the name, inert doublet) and the only Yukawa term where it appears is the one with the RH neutrinos. The $Z_2$ symmetry also forbids the normal Yukawa coupling involving the lepton doublets, RH neutrinos and the SM Higgs doublet. On the other hand, the allowed Yukawa coupling between the lepton doublets, RH neutrinos and the inert doublet does not lead to a Dirac-like mass term since the inert doublet does not take a VEV. As a result, light neutrino masses via Type-I seesaw is forbidden. However, the light neutrinos get mass radiatively at one-loop, where the RH neutrinos and the inert doublet run in the loop~\cite{Ma:2006km}.
The RH neutrinos protected by the $Z_2$ symmetry become the dark matter of the universe. The $Z_2$ symmetry allows the RH neutrinos to be coupled only to the Higgs sector and the $Z_{\mu\tau}$. 
We invoke a non-thermal production mechanism for the generation of DM in the early universe 
via the freeze-in mechanism \cite{Hall:2009bx} wherein the RH neutrinos are mainly produced 
by out-of-equilibrium decays of $Z_{\mu\tau}$ gauge bosons. 

The 3.5 keV $\gamma$ line can be explained by the decay of a heavy RH neutrino to another 
RH neutrino if the two states are nearly degenerate and the mass splitting is 3.5 keV 
\cite{Chiang:2014xra, Falkowski:2014sma}. 
Moreover the lifetime of the next-to-lightest neutrino has to be sufficiently long.
Both conditions are naturally realised in our scenario.
Indeed we will see that in the $U(1)_{L_{\mu} - L_{\tau}}$ symmetric limit, 
the $L_{\mu} - L_{\tau}$ symmetry enforces two completely degenerate states and 
one heavier state for the RH mass spectrum in our model. The two lighter degenerate RH 
neutrino states play the role of a two-component dark matter. The spontaneous breaking of 
$U(1)_{L_{\mu} - L_{\tau}}$ results in a small mass splitting between the two degenerate RH 
neutrinos, determined by the symmetry breaking scale and Yukawa couplings of the RH neutrinos. 
The lifetime of the heavier state is longer than the age of the Universe due both
to the phase-space suppression and to the small parameters needed to explain the light 
neutrino masses.

The rest of the article is organised in the following way. In Section \ref{model} we describe the
model in detail. In Section \ref{neutrino-part}, we discuss the effect of $\umt$ and its breaking 
on the RH neutrino mass spectrum.
In Section \ref{bee} we present our formalism for the freeze-in production of the RH neutrino DM. 
In section we show our results on the DM relic abundance and discuss all aspects related to it. 
In Section \ref{dm-result} and Section \ref{355-part} we will present our DM results
and explain the origin of 3.5
keV line in our model from the RH neutrino decay respectively.
We finally conclude in Section \ref{con}.
\section{Model}
\label{model}
The complete gauge group in our model is,
$\smgauge \times U(1)_{L_{\mu} - L_{\tau}}$. In addition to the SM particles, we augment our 
model with a SM scalar doublet, a SM scalar singlet and three RH neutrinos. We also impose a 
${\mathbb Z}_2$ symmetry to make the additional doublet inert. The ${\mathbb Z}_2$ charge of the 
RH neutrinos are also taken to be $-1$ to keep them stable, such that they could be dark matter 
candidates. The complete fermionic and scalar particle content
of the model and their corresponding charges under the different
symmetry groups are shown in Tables \ref{tab1} and \ref{tab2}:
\def\I{i}
\begin{center}
\begin{table}[h!]
\begin{tabular}{||c|c|c|c||}
\hline
\hline
\begin{tabular}{c}
    Gauge\\
    Group\\ 
    \hline
    
    ${\rm SU(2)}_{\rm L}$\\ 
    \hline
    ${\rm U(1)}_{\rm Y}$\\
    \hline
    $\mathbb{Z}_{2}$\\ 
\end{tabular}
&

\begin{tabular}{c|c|c}
    \multicolumn{3}{c}{Baryon Fields}\\ 
    \hline
    $Q_{L}^{i}=(u_{L}^{i},d_{L}^{i})^{T}$&$u_{R}^{i}$&$d_{R}^{i}$\\ 
    \hline
    $2$&$1$&$1$\\ 
    \hline
    $1/6$&$2/3$&$-1/3$\\
    \hline
    $+$&$+$&$+$\\ 
\end{tabular}
&
\begin{tabular}{c|c|c}
    \multicolumn{3}{c}{Lepton Fields}\\
    \hline
    $L_{L}^{i}=(\nu_{L}^{i},e_{L}^{i})^{T}$ & $e_{R}^{i}$ & $N_{R}^{i}$\\
    \hline
    $2$&$1$&$1$\\
    \hline
    $-1/2$&$-1$&$0$\\
    \hline
    $+$&$+$&$-$\\
\end{tabular}
&
\begin{tabular}{c|c|c}
    \multicolumn{3}{c}{Scalar Fields}\\
    \hline
    $\phi_{h}$&$\phi_{H}$&$\eta$\\
    \hline
    $2$&$1$&$2$\\
    \hline
    $1/2$&$0$&$1/2$\\
    \hline
    $+$&$+$&$-$\\
\end{tabular}\\
\hline
\hline
\end{tabular}
\caption{Particle contents and their corresponding
charges under SM gauge group and discrete group $\mathbb{Z}_2$.}
\label{tab1}
\end{table}
\end{center}
\begin{center}
\begin{table}[h!]
\begin{tabular}{||c|c|c|c||}
\hline
\hline
\begin{tabular}{c}
    Gauge\\
    Group\\ 
    \hline
    $\umt$\\ 
    
\end{tabular}
&
\begin{tabular}{c}
    \multicolumn{1}{c}{Baryonic Fields}\\ 
    \hline
    $(Q^{i}_{L}, u^{i}_{R}, d^{i}_{R})$\\ 
    \hline
    $0$ \\ 
    
\end{tabular}
&
\begin{tabular}{c|c|c}
    \multicolumn{3}{c}{Lepton Fields}\\ 
    \hline
    $(L_{L}^{e}, e_{R}, N_{R}^{e})$ & $(L_{L}^{\mu}, \mu_{R},
    N_{R}^{\mu})$ & $(L_{L}^{\tau}, \tau_{R}, N_{R}^{\tau})$\\ 
    \hline
    $0$ & $1$ & $-1$\\ 
    
\end{tabular}
&
\begin{tabular}{c|c|c}
    \multicolumn{3}{c}{Scalar Fields}\\
    \hline
    $\phi_{h}$ & $\phi_{H}$ & $\eta$ \\
    \hline
    $0$ & $1$ & $0$\\
\end{tabular}\\
\hline
\hline
\end{tabular}
\caption{Particle contents and their corresponding
charges under $\umt$.}
\label{tab2}
\end{table}
\end{center} 
The complete Lagrangian $\mathcal{L}$ for the present model
is as follows,
\begin{eqnarray}
\mathcal{L}&=&\mathcal{L}_{SM} + \mathcal{L}_{N} +
(D_{\mu}\phi_{H})^{\dagger} (D^{\mu}\phi_{H})
+(D_{\mu}\eta)^{\dagger} (D^{\mu}\eta) 
 + \sum_{j=\mu,\,\tau} Q^j\,\bar{L_j} \gamma_{\rho} L_j Z^{\rho}_{\mu\tau} \nn \\ 
&& 
-\frac{1}{4} {F_{\mu \tau}}_{\rho \sigma}
{F_{\mu \tau}}^{\rho \sigma}
- V(\phi_{h},\phi_{H},\eta)\,,
\label{lag}
\end{eqnarray}
where $\phi_h$ and $\eta$ are two SU(2)$_{L}$ doublets
while $\phi_H$ is a scalar singlet. Moreover,
$Q^j=1(-1)$ for $j=\mu(\tau)$ where
$L_j=\left(\nu_j\,\,\,\, j\right)^{T}$.
Here, one of the scalar doublets
namely $\eta$ which is odd under $\mathbb{Z}_2$ symmetry, 
does not have any Yukawa interaction
involving only SM fermions and acts like an inert doublet. For the same 
symmetry reason it 
does not have any VEV. The
field strength tensor for the extra neutral gauge
field $\zmt$ corresponding to
gauge group $\umt$ is denoted by $F_{\mu\tau}$.
In principle we should include a mixing term between the SM neutral 
gauge boson $Z$ and the new neutral gauge boson $Z_{\mu\tau}$. 
The experimental bound restricts this mixing to be $<10^{-3}$ 
br the LEP II \cite{Carena:2004xs, Appelquist:2002mw}.
In this work we assume no mixing between the neutral gauge
bosons of SM and $U(1)_{L_{\mu} - L_{\tau}}$.
Indeed, if such mixing is generated at the loop level,
we expect it to be suppressed not only by loop factors,
but also by the gauge coupling $ g_{\mu\tau}$\footnote{In this work,
to maintain the nonthermal nature of our DM candidates
we consider $\gmt\sim 10^{-11}$ (see Section \ref{dm-result}).}
rendering it negligible in our discussion.
The Lagrangian for the three RH neutrinos
$\mathcal{L}_{N}$ after obeying all the symmetry
has the following form,
\begin{eqnarray}
\mathcal{L}_{N}&=&
\sum_{i=e,\,\mu,\,\tau}\frac{i}{2}\bar{N_i}\gamma^{\mu}D_{\mu} N_{i} 
-\dfrac{1}{2}\,M_{ee}\,\bar{N_e^{c}}N_{e}
-\dfrac{1}{2}\,M_{\mu \tau}\,(\bar{N_{\mu}^{c}}N_{\tau}
+\bar{N_{\tau}^{c}}N_{\mu})  \nn \\ &&
-\dfrac{1}{2}\,h_{e \mu}(\bar{N_{e}^{c}}N_{\mu} 
+\bar{N_{\mu}^{c}}N_{e})\phi_H^\dagger
- \dfrac{1}{2}\,h_{e \tau}(\bar{N_{e}^{c}}N_{\tau} 
+ \bar{N_{\tau}^{c}}N_{e})\phi_H
\nn \\ &&
-\sum_{\alpha = e,\,\mu,\,\tau} h_{\alpha} \bar{L_{\alpha}}
\tilde {\eta} N_{\alpha} + h.c.\,,
\label{lagN}
\end{eqnarray}
where $\tilde{\eta} = i \sigma_{2} \eta^{*}$.
The potential $V(\phi_{h},\phi_{H},\eta)$ in Eq.~(\ref{lag})
contains all possible interaction terms involving the two SM scalar
doublets and one SM scalar singlet,
\begin{eqnarray}
V(\phi_h, \phi_H, \eta) &=& -\mu_{H}^{2}
\phi_{H}^{\dagger} \phi_{H} 
- \mu_{h}^{2} \phi_{h}^{\dagger} \phi_{h}
+ \mu_{\eta}^{2} \eta^{\dagger} \eta
+ \lambda_{1} (\phi_{h}^{\dagger} \phi_{h})^{2}
+ \lambda_{2} (\eta^{\dagger} \eta)^{2}  
+ \lambda_{3} (\phi_{H}^{\dagger} \phi_{H})^{2} \nn \\
&& 
+ \lambda_{12}(\phi_{h}^{\dagger} \phi_{h}) (\eta^{\dagger} \eta)
+ \lambda_{13}(\phi_{h}^{\dagger} \phi_{h}) (\phi_{H}^{\dagger} \phi_{H})
+ \lambda_{23} (\phi_{H}^{\dagger} \phi_{H}) (\eta^{\dagger} \eta)
+ \lambda_{4} (\phi_{h}^{\dagger} \eta) (\eta^{\dagger} \phi_{h})\nn \\ 
&&+ \frac{1}{2} \lambda_{5} \left( (\phi_{h}^{\dagger} \eta)^{2}
+ h.c.\right)\,.
\label{int}
\end{eqnarray}
After spontaneous breaking of $\umt$ and $\smgauge$,
 the scalars take the 
following form,
\begin{eqnarray}
\phi_{h}=
\begin{pmatrix}
0 \\
\dfrac{v+H}{\sqrt{2}}
\end{pmatrix}\,,
\,\,\,\,
\phi_{H}=
\begin{pmatrix}
\dfrac{v_{\mu \tau} + H_{\mu \tau}}{\sqrt{2}}
\end{pmatrix}\,,\,\,\,\,
\eta = 
\begin{pmatrix}
\eta^{+} \\
\dfrac{\eta^{0}_R + i\,\eta^{0}_I}{\sqrt{2}}
\end{pmatrix}
\,.
\label{phih}
\end{eqnarray} 
There is mixing between the neutral components of
$\phi_h$ and $\phi_H$, and the off diagonal elements of the mass matrix are proportional
to the parameter $\lambda_{13}$. After diagonalising the mass 
matrix one obtains two physical scalar states $h_1$ and $h_2$. 
Masses of $h_1$, $h_2$ and mixing angle $\alpha$ are given by
\begin{eqnarray}
M^2_{h_1} &=& \lambda_1 v^2 + \lambda_3 \vmt^2 -
\sqrt{(\lambda_3 \vmt^2 - \lambda_1 v^2)^2 +
(\lambda_{13}\,v\,\vmt)^2}\ ,
\label{massh1} \\
M^2_{h_2} &=& \lambda_1 v^2 + \lambda_3 \vmt^2 + 
\sqrt{(\lambda_3 \vmt^2 - \lambda_1 v^2)^2 +
(\lambda_{13}\,v\,\vmt)^2}\,,
\label{massh2}\\
\tan 2\alpha &=& \dfrac{\lambda_{13}\,\vmt\,v}
{\lambda_3 \vmt^2 - \lambda_1 v^2}\,.
\label{scalarmix}
\end{eqnarray} 
The lighter Higgs state $ h_1 $, for small mixing angle $ \alpha $ and $ \vmt \gg v$, 
behaves as the Standard Model Higgs observed at the LHC~\cite{Aad:2012tfa,Chatrchyan:2012xdj}
and therefore we will take its mass to be $125.5 $ GeV.
From the above Eq.\,(\ref{massh1})-(\ref{scalarmix}),
we can also write down the quartic couplings in terms of the physical masses
of the Higgses $M_{h_1}$ and $M_{h_2}$ and the mixing angle $\alpha$. The expressions
are as follows,
\begin{eqnarray}
\lambda_{3} &=& \dfrac{M_{h_{1}}^{2} + M_{h_{2}}^{2} +
(M_{h_{2}}^{2} - M_{h_{1}}^{2})\cos 2 \alpha}{4\,v_{\mu\tau}^{2}}\,,\nn \\
\lambda_{1}&=& \dfrac{M_{h_{1}}^{2} + M_{h_{2}}^{2} +
(M_{h_{1}}^{2} - M_{h_{2}}^{2})\cos 2 \alpha}{4\,v^{2}}\,,\nn\\
\lambda_{13} &=& \dfrac{(M_{h_{2}}^{2}-M_{h_{1}}^{2})
\cos \alpha \sin \alpha}{v\,v_{\mu\tau}}\,,
\label{quartic}
\end{eqnarray}
In order to obtained a stable {\it vacuum}, the quartic couplings need to
satisfy the following inequalities, 
\begin{eqnarray}
&&\lambda_1 \geq 0, \lambda_2 \geq 0, \lambda_{3} \geq 0,\nonumber \\
&&\lambda_{12} \geq - 2\sqrt{\lambda_1\,\lambda_2},\nonumber \\
&&\lambda_{13} \geq - 2\sqrt{\lambda_1\,\lambda_3},\nonumber \\
&&\lambda_{23} \geq - 2\sqrt{\lambda_2\,\lambda_3},\nonumber \\
&&\lambda_{12} + \lambda_{4} - |\lambda_{5}| \geq - 2\sqrt{\lambda_1\,\lambda_{2}},\nonumber \\
&&\sqrt{\lambda_{13}+2\sqrt{\lambda_1\,\lambda_3}}\sqrt{\lambda_{23}+
2\sqrt{\lambda_2\,\lambda_{3}}}
\sqrt{\lambda_{12} + \lambda_{4} - |\lambda_{5}|+
2\sqrt{\lambda_1\,\lambda_{2}}} \nonumber \\ 
&&+ 2\,\sqrt{\lambda_1 \lambda_2 \lambda_{3}} + \lambda_{13} \sqrt{\lambda_{2}}
+ \lambda_{23} \sqrt{\lambda_1} + (\lambda_{12} + \lambda_{4} - |\lambda_{5}|)
\sqrt{\lambda_3} \geq 0 \,\,\,\,,\nn \\
&&\sqrt{\lambda_{13}+2\sqrt{\lambda_1\,\lambda_3}}\sqrt{\lambda_{23}+
2\sqrt{\lambda_2\,\lambda_{3}}}
\sqrt{\lambda_{12} +
2\sqrt{\lambda_1\,\lambda_{2}}} \nonumber \\ 
&&+ 2\,\sqrt{\lambda_1 \lambda_2 \lambda_{3}} + \lambda_{13} \sqrt{\lambda_{2}}
+ \lambda_{23} \sqrt{\lambda_1} + \lambda_{12}
\sqrt{\lambda_3} \geq 0 \,\,\,\,.
\label{ineuality}
\end{eqnarray}
As we will see in the result section (Section \ref{dm-result}), in our analysis 
the value of the extra singlet scalar {\it vev} is around $10^{14}$ GeV, mass of BSM
Higgs $M_{h_2} = 5$ TeV and the mixing angle between the neutral components of
Higgses $\alpha$ = 0.01. Hence, we get the following values for the quartic couplings
by using the Eq.\,(\ref{quartic}),
\begin{eqnarray}
\lambda_1 = 0.15,\,\,\,\lambda_3 = 1.25 \times 10^{-21}\,\, {\rm and}\,\, \lambda_{13}
= 1.01 \times 10^{-11}\,.
\end{eqnarray}
All the values of the quartic couplings as shown above
are positive and in the present case the quartic couplings which are
related to the inert doublet are free parameters (except $\lambda_5$,
which we have considered $\sim 10^{-3}$ to obtain light
neutrino masses in sub-eV range),
hence all the inequalities as prescribed in
Eq.\,(\ref{ineuality}) are inevitably satisfied.

On the other hand the masses of the inert doublet
components after symmetry breaking can be
expressed in the following form,
\begin{eqnarray}
M^2_{\eta^\pm} &=&
\mu^2_{\eta}+\dfrac{1}{2}\left(\lambda_{12} v^{2} +
\lambda_{23} \vmt^{2}\right)\,, \nn \\
M^{2}_{\eta^{0}_R} &=&
\mu^2_{\eta} + \dfrac{1}{2}\lambda_{23}\,
\vmt^{2} + \dfrac{1}{2}(\lambda_{12} + \lambda_{4}
+ \lambda_{5})\,v^{2}\,, \nn \\
M^{2}_{\eta^0_I} &=&
\mu^2_{\eta} + \dfrac{1}{2}\lambda_{23}\,\vmt^{2}
+ \dfrac{1}{2}(\lambda_{12}+ \lambda_{4} - \lambda_{5})\,v^{2}\,,
\end{eqnarray}
The mass term for the extra neutral gauge boson $\zmt$
is also generated  when $\phi_{H}$ acquires a
nonzero VEV $\vmt$ such that 
\begin{eqnarray}
\mzmt = \gmt\,\vmt\,,
\end{eqnarray} 
where $\gmt$ is the gauge coupling corresponding to gauge group
$\umt$. 
In this model all three RH neutrinos are odd under the
$\mathbb{Z}_2$ symmetry. 
However, the mass of $N_1$ comes out to be 
higher than that of $N_2$ and $N_3$, so that
$N_1$ can decay to the lighter RH neutrinos. Also,
we will see in Section \ref{neutrino-part} that the masses
of $N_2$ and $N_3$ are nearly degenerate because of the 
$L_{\mu}-L_{\tau}$ symmetry, so that both can
play the role of dark matter candidate. 
Furthermore, in Section \ref{bee} we will show that the RH neutrinos 
can be produced by the freeze-in mechanism in the early Universe, 
which requires a tiny gauge coupling $\gmt\sim \mathcal{O}(10^{-11})$.
Thus, in order to have a TeV scale gauge boson $\zmt$
we need large $\vmt$. Therefore, by choosing appropriate
values of the relevant model parameters we can make the masses of
inert doublet components higher than the reheat temperature
of the universe so that their effect on the production
of $N_2$ and $N_3$ can be safely neglected.
\section{Heavy and Light Neutrino Masses}
\label{neutrino-part}
In this section we will show how the $\umt$ symmetry determines the 
mass spectrum and mixing angles of all the six neutrinos, the three heavy ones as well as the 
three light ones. The relevant part of the Lagrangian was given in Eq.~(\ref{lagN}) where the first term gives the kinetic part while the rest give the mass terms and Yukawa terms involving the neutrinos. After $\umt$ and electroweak symmetry breaking the mass matrix for the RH neutrinos is given by
\begin{eqnarray}
\mathcal{M}_{R} = \left(\begin{array}{ccc}
M_{ee} ~~&~~ \dfrac{ \vmt}{\sqrt{2}} h_{e \mu}
~~&~~\dfrac{\vmt}{\sqrt{2}} h_{e \tau} \\
~~&~~\\
\dfrac{\vmt}{\sqrt{2}} h_{e \mu} ~~&~~ 0
~~&~~ M_{\mu \tau} \,e^{i\xi}\\
~~&~~\\
\dfrac{\vmt}{\sqrt{2}} h_{e \tau} ~~&
~~ M_{\mu \tau}\,e^{i\xi} ~~&~~ 0 \\
\end{array}\right) \,,
\label{mncomplex}
\end{eqnarray}
where the terms involving the VEV $v_{\mu\tau}$ appear after $\umt$ breaking.
In the limit that $\umt$ is unbroken, the RH neutrino mass matrix is given by
\begin{eqnarray}
\mathcal{M}_{R} = \left(\begin{array}{ccc}
M_{ee} ~~&~~ 0 
~~&~~0 \\
~~&~~\\
0 ~~&~~ 0
~~&~~ M_{\mu \tau} \,e^{i\xi}\\
~~&~~\\
0 ~~&
~~ M_{\mu \tau}\,e^{i\xi} ~~&~~ 0 \\
\end{array}\right) \,.
\label{mncomplex1}
\end{eqnarray}
Eigenvalues of Eq.~(\ref{mncomplex1}) are
\begin{eqnarray}
M'_{2/3} &=& \pm M_{\mu\tau}e^{i\xi} \nn \\
M'_{1} &=& M_{ee}
\,,
\end{eqnarray}
giving very naturally two degenerate RH neutrinos with opposite parity. The $\umt$ breaking terms in Eq.~(\ref{mncomplex}) brings corrections to the RH neutrino mass spectrum, breaking the degeneracy between $N_2$ and $N_3$.
The mass splitting between them is given at first order for $ M_{ee} \gg M_{\mu\tau}$ by
\begin{eqnarray}
\Delta M_{23} =  \frac{ (h_{e \mu} + h_{e \tau})^2 \vmt^2}{2 M_{ee}}
\,.
\end{eqnarray}
Hence, the mass splitting between $N_2$ and $N_3$ depends on the $\umt$ breaking VEV $v_{\mu\tau}$ and the Yukawa couplings $h_{e\mu}$ and $h_{e\tau}$. In what follows, we will see that $v_{\mu\tau}$ will be determined by the choice of the $Z_{\mu\tau}$ gauge boson. However, the Yukawa couplings $h_{e\mu}$ and $h_{e\tau}$ can be suitably adjusted to yield a mass splitting 
of 3.5 keV, needed to explain the 3.5 keV X-ray line from $N_2\to N_3\gamma$ decay. 

\begin{figure}[h!]
\centering
\includegraphics[angle=0,height=3.5cm,width=8.0cm]{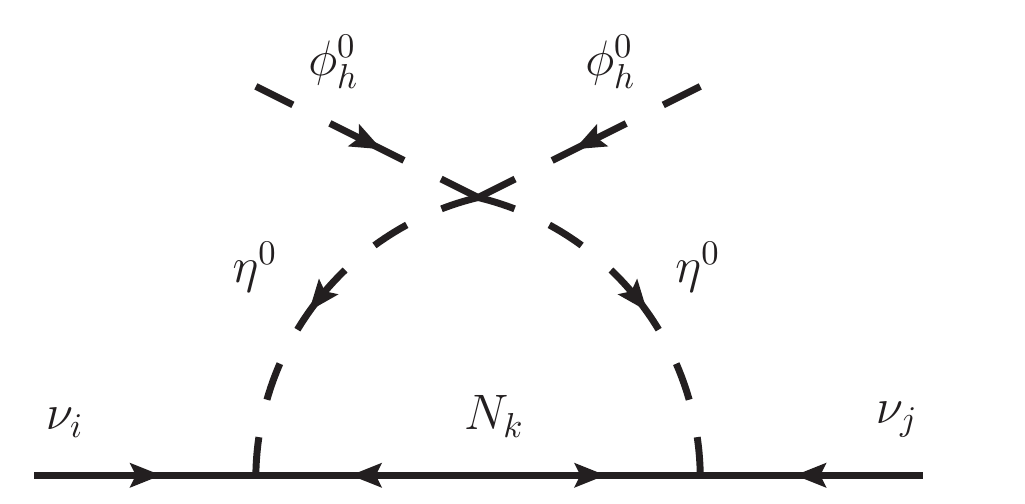}
\caption{Radiative neutrino mass generation by one loop.}      
\label{a1}
\end{figure}

Despite having the RH neutrinos in this model, the masses for light neutrinos cannot be generated by the Type-I seesaw mechanism since the normal Yukawa term involving the RH neutrinos, lepton doublets and the standard model Higgs $\phi_h$ is forbidden by the ${\mathbb Z_2}$ symmetry. The other Yukawa term between the RH neutrinos, lepton doublets and inert doublet $\eta$ is allowed, but $\eta$ does not take any VEV. Hence, there is no mass term for the light neutrinos at the tree-level. However, masses for the light neutrinos gets generated radiatively at the one-loop level \cite{Ma:2006km} through the diagram shown in Fig.~\ref{a1},  
giving the following mass matrix for the light neutrinos \cite{Ma:2006km}
\begin{eqnarray}
M^{\nu}_{ij} = \sum_k \frac{y_{ik}\,y_{jk}\,M_{k}}{16\,\pi^2}
\left[\frac{M^2_{\eta_R^0}}{M^2_{\eta_R^0}-M^2_k}\,
\ln \frac{M^2_{\eta_R^0}}{M^2_k}-
\frac{M^2_{\eta_I^0}}{M^2_{\eta_I^0}-M^2_k}
\ln \frac{M^2_{\eta_I^0}}{M^2_k}\,\right]\,,
\end{eqnarray}  
where $M_k$ is the mass of $k^{\rm th}$ RH neutrino while
$M_{\eta^0_{R},\,\eta^0_{I}}$ is the mass of $\eta^0_{R,\,I}$. The 
quantities $y_{j i} = h_{j} U_{j i}$, where $h_j$ are the Yukawa 
couplings in the last term of Eq.~(\ref{lagN}) and $U_{ji}$ are the 
elements of the RH neutrino mixing matrix
since the flavour basis ($N_{\alpha}$,
$\alpha=1,\,2,\,3$) of the RH neutrinos and their mass basis ($N_i$, $i=1,\,2\,3$) 
are related by a unitary
transformation, $N_{\alpha} = \sum U_{\alpha i} N_{i}$.
If we put this relation into the last term of Eq.~(\ref{lagN}), one can
write the Yukawa term involving SM leptons and RH neutrinos
in the following way
\begin{eqnarray}
\mathcal{L}_{N} \supset  h_{j} \bar{L_{j}}
\tilde {\eta} U_{j i} N_{i}\, = y_{ji} \bar{L_{j}}
\tilde {\eta} N_{i}\,.
\label{yuk}
\end{eqnarray}
If we consider the mass square difference between $\eta^0_{R}$
and $\eta^0_{I}$ i.e. $M^2_{\eta^0_{R}}-M^2_{\eta^0_{I}}
= \lambda_5 v^2 << M^2_0$ where $M^2_0 = (M^2_{\eta^0_{R}}
+M^2_{\eta^0_{I}})/2$ then the above expression reduces to
the following form,
\begin{eqnarray}
M^{\nu}_{ij} = \frac{\lambda_{5} v^{2}}{16 \pi^{2}}
\sum_{k} \frac{y_{ik}\,y_{jk}\,M_{k}}{M_{0}^{2} - M_{k}^{2}}
\left[ 1 - \frac{M_{k}^{2}}{M_{0}^{2} - M_{k}^{2}}\, \ln
\frac{M_{0}^{2}}{M_{k}^{2}} \right]\,.
\end{eqnarray}
In this work we have considered the masses of
inert scalars greater than the reheat temperature of
the Universe, i.e. $M_{\eta^0_{R,\,I}}\sim\,10^{6}$ GeV. The masses
of RH neutrinos we consider to be around $\sim 100$
GeV. If we take the parameter $\lambda_{5} \sim 10^{-3}$
and $v = 246$ GeV, 
then to obtain the neutrino masses of the order of
$M_{\nu} \sim 10^{-11}$ GeV, we need $y_{ji}^2 \sim 10^{-1}$
which can be easily obtained. The $\umt$ breaking ensures that the 
mixing angle $\theta_{13}$ is non-zero and $\theta_{23}$ is non-maximal. 

\section{Production of Dark Matter}
\label{bee}

We consider the non-thermal production of dark matter candidates. Hence, the initial number densities of these particles are assumed to be negligibly small and their interactions with the particles in the thermal bath are also extremely feeble. As mentioned before, the lighter RH neutrino states $N_2$ and $N_3$ are our dark matter candidates, stabilised by the $Z_2$ symmetry. Because of their gauge and $Z_2$ charges they could be produced only through the decay of $Z_{\mu\tau}$ and $h_1$ \footnote{Since the mass of the SM-like Higgs has to be kept at 125.5 GeV, the decay channel $h_1\to N_i N_j$ will be kinematically allowed only for lighter $N_i/N_j$ masses.} and $h_2$ bosons. In what follows, we will see that the dominant production channel for the RH neutrinos is via the decay of $Z_{\mu\tau}$. In order for the total abundances of $N_2$, $N_3$ to match the observed DM relic density at the present epoch, the gauge coupling has to be small $\gmt\la 10^{-11}$.
 Since all the interactions of $\zmt$ are proportional to the gauge coupling $\gmt$, the requirement of such a tiny gauge coupling makes the additional neutral gauge boson $\zmt$ also decoupled from the thermal bath. Therefore, before computing the DM number density we first need to know the distribution function of mother particle $\zmt$ by solving the relevant Boltzmann equation. 
The most general form of the Boltzmann equation describing the distribution function of any 
species can be expressed as,
\begin{eqnarray}
\hat{L}\,[f] = \mathcal{C}\,[f]
\end{eqnarray} 
where $\hat{L}$ is the Liouville operator and $f$
is the distribution function which we want to compute
while in the RHS the term $\mathcal{C}$ contains 
interaction processes which are responsible for changing
the number density of the species under considering.
$\mathcal{C}$ is known as the collision term.  
If one considers an isotropic and homogeneous
Universe then using the FRW metric, the Liouville operator
\footnote{General form of the Liouville operator is,
$\hat{L} = p^{\alpha} \frac{\partial}{\partial x^{\alpha}} -
\Gamma^{\alpha}_{\beta \gamma} p^{\beta} p^{\gamma}
\frac{\partial}{\partial p^{\alpha}}$ where $p^{\alpha}$
is the four momentum and $\Gamma^{\alpha}_{\beta \gamma}$
is the affine connection by which gravitational
interaction enters in the equation.} takes the following form,
\begin{eqnarray}
\hat{L} = \frac{\partial}{\partial t} -
H\, p\, \frac{\partial}{\partial p}\,,
\label{liouville}
\end{eqnarray}
where $p$ is magnitude of three momentum and $H$ is the
Hubble parameter. Now, we change the variables $(p,\,t)$
to a new set of variables ($\xi_p$,\,$r$) using a
transformation as mentioned in Ref. \cite{Konig:2016dzg}
\begin{eqnarray}
r = \frac{M_{sc}}{T},\,\,\,\, \xi_{p} =
\left(\frac{g_{s} (T_{0})}{g_{s}(T)} \right)^{1/3}
\frac{p}{T}\,.
\end{eqnarray}
$M_{sc}$ is some reference mass scale.\,\,Using
the time-Temperature relationship 
$\frac{dT}{dt}=-H\,T\,\left(1+\frac{T\,g^{\prime}_s(T)}{3\,g_s(T)}
\right)^{-1}$, the Liouville operator defined in Eq.~(\ref{liouville})
can be reduced to the following form containing a
derivative with respect to a single variable, i.e. 
\begin{eqnarray}
\hat{L} = r\,H \left( 1 + \frac{T g_{s}^{\prime}}
{3 g_{s}} \right)^{-1}\frac{\partial}{\partial\,r}
\end{eqnarray}
where $g_{s}(T)$ and $g_{s}^{\prime}(T)$ are the effective
number of degrees of freedom (d.o.f) related to entropy of the
Universe and its derivative with respect to the temperature $T$.

The Boltzmann equation to determine the distribution
function ($f_{\zmt}$) of $\zmt$ is then given by,
\begin{eqnarray}
\hat{L} f_{Z_{\mu\tau}} = \sum_{i = 1,2}
\mathcal{C}^{h_{i} \rightarrow Z_{\mu\tau} Z_{\mu\tau}}
+ \mathcal{C}^{Z_{\mu\tau} \rightarrow\,\, {all}}\,,
\label{zmt_prod}
\end{eqnarray}
where the first term in the RHS represents the production
of $\zmt$ from the decays of scalars $h_1$ and $h_2$ while
the second term describing the depletion of $\zmt$ due
to its all possible decay modes. The expressions of
collision terms 
$\mathcal{C}^{h_{i} \rightarrow Z_{\mu\tau} Z_{\mu\tau}}$
and $\mathcal{C}^{Z_{\mu\tau} \rightarrow\,\, {all}}$
are given in Appendix\,\,\,\ref{App:AppendixA}.
Note that generically also scattering processes, which change
the $ \zmt $ number, are present, but those give a subleading contribution
compared to the decay (see e.g. the Appendix of \cite{Arcadi:2014dca}
for a discussion).

Once we numerically evaluate the non thermal momentum distribution
of the gauge boson $Z_{\mu \tau}$, we can easily determine
the number density of $Z_{\mu\tau}$ using following relation
\begin{eqnarray} 
n_{Z_{\mu\tau}}(r) &=& \dfrac{g\, T^3}{2\pi^2} \,
\mathcal{B}(r)^3 \int d\xi_p\,\xi_p^2\, f_{Z_{\mu\tau}}(\xi_p,\,r)\,,
\label{n}
\end{eqnarray}
where
\begin{eqnarray}
\mathcal{B}(r) = \left(\frac{g_{s}
(T_{0})}{g_{s}(T)} \right)^{1/3} = \left(\frac{g_{s} (M_{sc}/r)}
{g_{s}(M_{sc}/r_0)} \right)^{1/3}\,.
\label{convfact}
\end{eqnarray}
Here $T_0$ is the initial temperature and $M_{sc}$ is
some reference mass scale. In this work we take
$T_0 = 10$ TeV and $M_{sc} = M_{h_1} = 125.5$ GeV, the
mass of SM Higgs boson. The entropy density
of the Universe is given by \cite{Kolb:1990vq},
\begin{eqnarray}
s &=& \dfrac{2\pi^2}{45} \, g_s(T)\,T^3\,.
\label{s}
\end{eqnarray}
Therefore, after determining the number density of $Z_{\mu\tau}$
and the entropy of the Universe one can determine the comoving
number density using the following relation,
\begin{eqnarray}
Y_{Z_{\mu\tau}} = \frac{n_{Z_{\mu\tau}}}{s} \,.
\end{eqnarray} 

Finally, to determine the comoving number densities
of DM components $N_2$ and $N_3$, we need to solve the
relevant Boltzmann equation for $N_2$ and $N_3$, which
can be written in a generic form,
\begin{eqnarray}
\frac{d Y_{N_j}}{d r} &=& \frac{V_{ij}\, M_{pl}\, r\, \sqrt{g_{\star} (r)}}
{1.66\, M_{sc}^{2}\, g_{s}(r)} \left[ \sum_{k=1,2}\sum_{i = 1,2,3}
\langle \Gamma_{h_k \rightarrow N_{j}\,N_{i}}\rangle
(Y_{h_k} - Y_{N_j} Y_{N_i})\,\right] \nonumber \\ 
&&+ \, \frac{V_{ij}\, M_{pl}\, r\, \sqrt{g_{\star} (r)}}
{1.66\, M_{sc}^{2}\, g_{s}(r)} \sum_{i=1,2,3} \langle 
\Gamma_{Z_{\mu\tau} \rightarrow N_{j} N_{i}} \rangle_{NTH}\,
(Y_{Z_{\mu\tau}} - Y_{N_j} Y_{N_i})\,,
\label{be}
\end{eqnarray}  
where $M_{pl}$ is the Planck mass while $g_{\star}(r) =
\frac{g_s(r)}{\sqrt{g_{\rho}(r)}}\left(1-\frac{1}{3}
\frac{d\,\ln g_s(r)}{d\ln r}\right)$ is
a function of $g_{\rho}(r)$ and $g_{s}(r)$. 
The parameter $V_{ij} = 2$ for $i=j$ and equal to 1 otherwise.
The first term in the above equation represents the production
of $N_j$ from the decays of scalar fields $h_1$ and $h_2$.
Since these scalar fields remain in thermal equilibrium
throughout their cosmological evolution, one can
consider their distribution function as
Maxwell-Boltzmann distribution. Therefore the thermal
averaged decay width for a process $h_(k)\rightarrow N_j\,N_i$
is given by~\cite{Gondolo:1990dk} 
\begin{eqnarray}
\langle \Gamma_{h_{k} \rightarrow N_{j}\, N_{i}} \rangle = 
\Gamma_{h_{k} \rightarrow N_{j}\, N_{i}}\,
\frac{K_{1}\left(r\,\frac{M_{h_k}}{M_{sc}}\right)}
{K_{2}\left(r\,\frac{M_{h_k}}{M_{sc}}\right)}\,,
\label{zdk}
\end{eqnarray}
\begin{figure}[h!]
\centering
\includegraphics[angle=0,height=9cm,width=10cm]{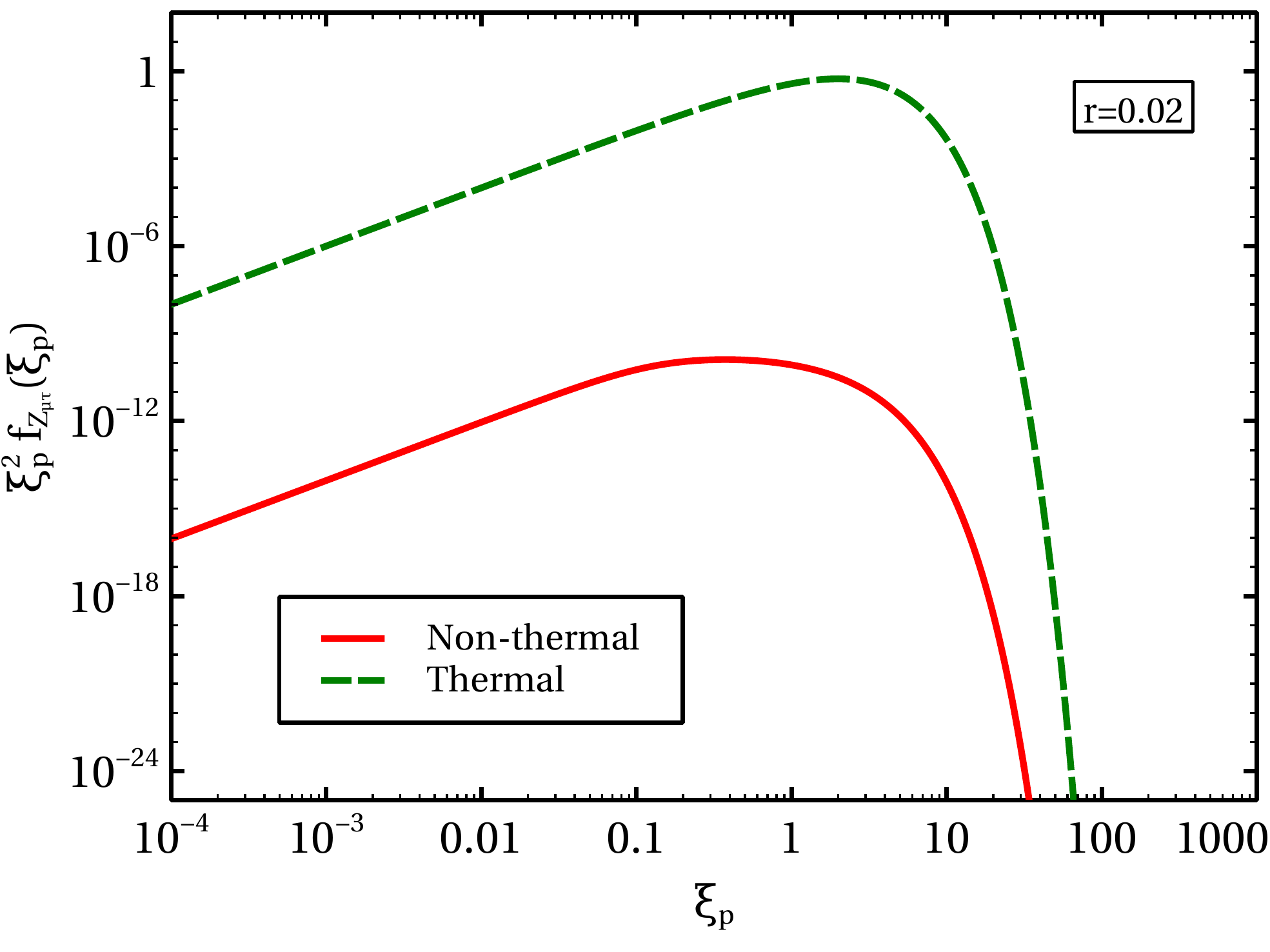}
\caption{Thernal and Non-thermal distribution function
of $Z_{\mu\tau}$ gauge boson.}
\label{dis-fz}
\end{figure}
where $K_{i}$ is the Modified Bessel function of $i^{th}$ kind.
As the neutral gauge boson $\zmt$ is not in
thermal equilibrium (due to very small value of $\gmt$),
one cannot assume a Maxwell-Boltzmann distribution function
for $\zmt$. The distribution $f_{\zmt}$ of $\zmt$ can be
found by solving Eq.~(\ref{zmt_prod}) and we have shown it in
Fig.\,\ref{dis-fz}. Although the shape of the distribution is
similar in both cases but they differ by magnitude because in the current case 
$Z_{\mu\tau}$ is always out of equilibrium and
never attains equilibrium value. Once we get the
distribution function $f_{\zmt}$ the non-thermal average
of decay width for the process $Z_{\mu\tau} \rightarrow N_{j} N_{i}$
can be computed as follows
\begin{eqnarray}
\langle \Gamma_{Z_{\mu\tau} \rightarrow N_{j} N_{i}}
\rangle_{NTH} = M_{Z_{\mu\tau}} 
\Gamma_{Z_{\mu\tau} \rightarrow N_{j} N_{i}}
\frac{\int \frac{f_{Z_{\mu\tau}}(p)}{\sqrt{p^{2}
+M_{Z_{\mu\tau}}^{2}}} d^{3} p}
{\int f_{Z_{\mu\tau}}(p) d^{3} p}\,.
\label{hdk}
\end{eqnarray}
All the relevant decay widths of $h_2$ and $Z_{\mu\tau}$
needed in Eq.~(\ref{be}) are given in Appendix\,\ref{App:AppendixA}
in detail. After solving the above Boltzmann equations
for j=2 and j=3, we can determine the comoving number density
of the DM candidates $N_2$ and $N_3$. Therefore, one can
easily determine the total DM relic density
for $N_2$ and $N_3$ candidates
by using the following relation~\cite{Edsjo:1997bg},
\begin{eqnarray}
\Omega_{DM}h^2 &=& 2.755\times 10^8
\bigg(\dfrac{M_{N_2}}{\rm GeV}\bigg)
\,Y_{N_2}(T_{\rm Now}) + 2.755\times 10^8
\bigg(\dfrac{M_{N_3}}{\rm GeV}\bigg)
\,Y_{N_3}(T_{\rm Now})\,. 
\label{relic}
\end{eqnarray}

\section{Results}
\label{dm-result}

\begin{figure}[h!]
\centering
\includegraphics[angle=0,height=7cm,width=8cm]{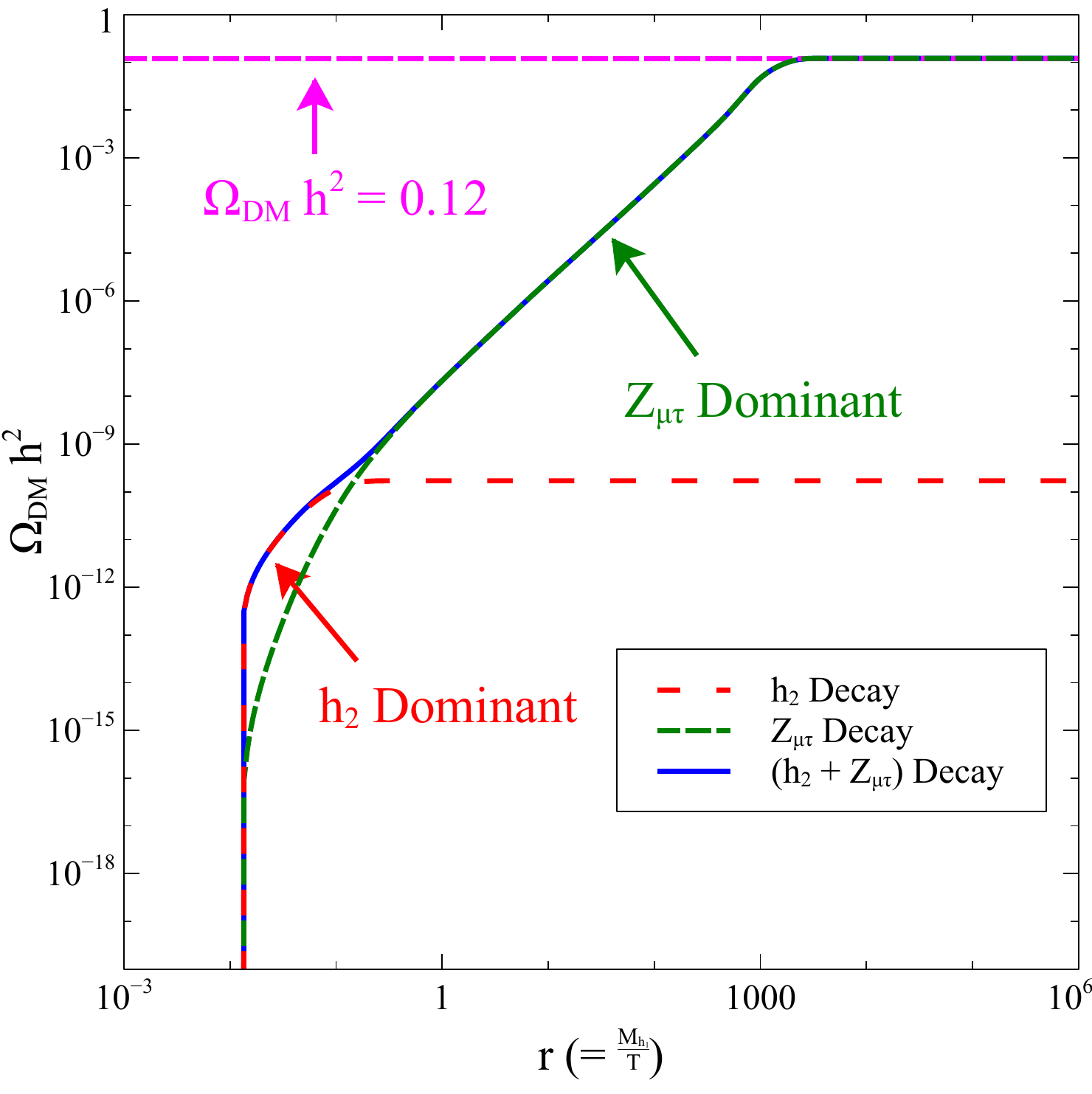}
\includegraphics[angle=0,height=7cm,width=8cm]{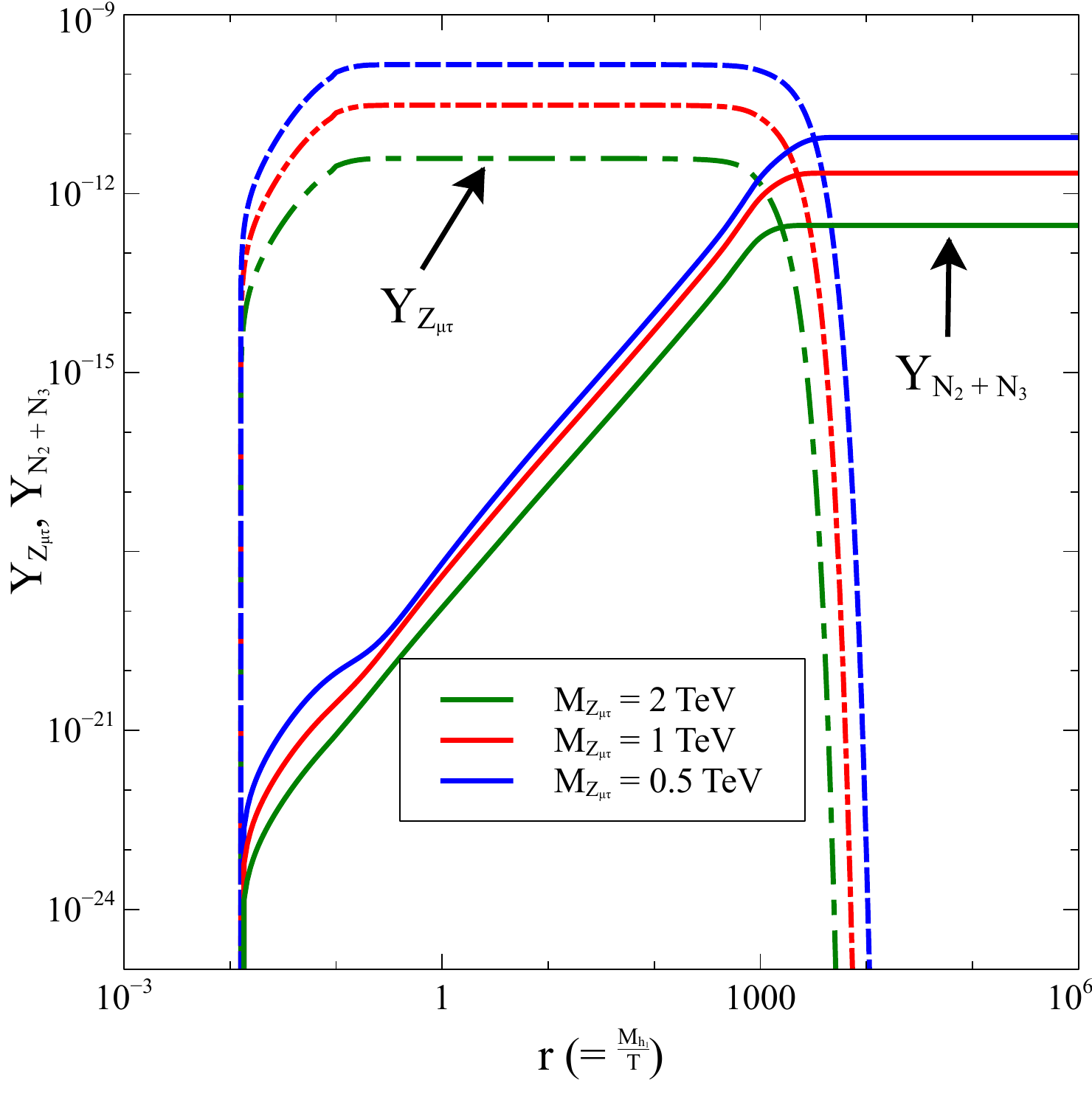}
\caption{Left panel: Variation of relic density with $r$ and
contributions from $h_2$ and $\zmt$ in the DM production. Right panel: Variation
of comoving number density of $\zmt$ and $N_2$, $N_3$ with $r$
for three different values of gauge boson mass.
Other parameters
have been kept fixed at $g_{\mu\tau} = 1.01 \times 10^{-11}$,
mixing angle $\alpha = 0.01$, gauge boson mass $M_{\zmt} = 1$ TeV, DM mass $M_{DM} = 100$ GeV,
BSM Higgs mass $M_{h_2} = 5$ TeV and RH neutrinos masses
$M_{N_1}$ = 150 GeV and $M_{DM}=M_{N_2} \simeq M_{N_3}$ = 100 GeV.}
\label{3a}
\end{figure}

Using Eqs.~(\ref{be}), (\ref{zdk}), (\ref{hdk}) and (\ref{relic}) we numerically compute 
the DM abundance. In the left panel of Fig.~\ref{3a} we show the time evolution of the DM relic 
density with $r(=M_{h_1}/T)$. The left panel of the this figure shows the comparative 
contribution for the two DM production channels, $Z_{\mu\tau}\to N_iN_j$ and $h_2\to  N_iN_j$. 
We have taken masses of the RH neutrinos $N_2$ and $N_3$ as 100 GeV and hence the decay of 
SM-like Higgs $h_1$ to a pair of RH neutrinos is kinematically forbidden. 
From the left panel we see that for the large value of BSM Higgs mass ($M_{h_2}\sim 5$ TeV), 
the DM production at low $r$ (which corresponds to high $T$) is dominated by $h_2$ decay. 
However, as the temperature of the universe falls and goes below the mass of the $Z_{\mu\tau}$ gauge bosons, 
they get produced, and for high value of $r$ (which corresponds to comparably lower temperature of the universe), 
the DM production via the $Z_{\mu\tau}$ decay channel dominates. The reason for this dominance can be understood as follows. 
From Eqs.~(\ref{dkz}) and (\ref{dkh}) given in Appendix A, we see that the decay width $\Gamma_{Z_{\mu\tau}\to N_iN_j} \propto M_{Z_{\mu\tau}}g_{\mu\tau}^2$ while $\Gamma_{h_2 \to N_iN_j} \propto M_{h_2} h_{e\alpha}h_{e\beta}$, where $h_{e\alpha}h_{e\beta}$ are products of two any of the Yukawa couplings $h_{e\mu}$ and $h_{e\tau}$ that appeared in Eq.~(\ref{lagN}). Since we have chosen $M_{Z_{\mu\tau}} \sim M_{h_2}$ we can write
\begin{eqnarray}
\frac{\Gamma_{Z_{\mu\tau}\to N_iN_j}}{\Gamma_{h_2 \to N_iN_j} } \propto \frac{g_{\mu\tau}^2}{h_{e\alpha}h_{e\beta}}
\,,
\label{fraction1}
\end{eqnarray}
Since the Yukawa couplings $h_{e\alpha}$ appear as the $\umt$ breaking terms in the RH neutrino mass matrix which instruments the splitting of 3.5 keV between $N_2$ and $N_3$ we have from Eq.~(\ref{mncomplex}) 
\begin{eqnarray}
V_{e\alpha} = \frac{h_{e\alpha} v_{\mu\tau}}{\sqrt{2}} \sim 0.1~{\rm GeV}
\,. 
\end{eqnarray}
Inserting this in Eq.~(\ref{fraction1}) and using the relation $M_{Z_{\mu\tau}} = g_{\mu\tau}v_{\mu\tau}$ we get
\begin{eqnarray}
\frac{\Gamma_{Z_{\mu\tau}\to N_iN_j}}{\Gamma_{h_2 \to N_iN_j} } \propto \frac{M_{Z_{\mu\tau}}^2}{V_{e\alpha}^2}
\,,
\label{fraction2}
\end{eqnarray}
explaining the dominance of the $Z_{\mu\tau}$ decay channel. 

In the right panel of Fig.~\ref{3a} we show the variation
of the comoving number densities of the $Z_{\mu\tau}$ gauge boson {\it vis-a-vis} that of the sum of
$N_2$ and $N_3$. We show this as function of $r$ for three different values of 
the gauge boson mass $M_{\zmt}$.

The abundance $Y_{\zmt}$ (indicated by the dash line) has an initial rise, then flattens
and finally decays. One can see from Eq.~(\ref{zmt_prod}) that there are two
collision terms in the Boltzmann Equation, one for $Z_{\mu\tau}$ production and 
another one for its decay to all possible channels and they are active at different times. 
Note that the maximal abundance of $Z_{\mu\tau}$ can be easily estimated also by the
analytic formula for FIMP production, i.e. for $ M_{\zmt} \ll M_{h_2} $
\be
\Omega^{FI} h^2 = 1.09 \times 10^{27} \frac{g}{g_S^{3/2}} \frac{M_{\zmt}}{M_{h_2}^2}  
\Gamma_{h_2 \to Z_{\mu\tau} Z_{\mu\tau}} \sim
2.18 \times 10^{24} \frac{ g_{\mu\tau}^2 M_{h_2}}{32 \pi M_{\zmt}} = 8.54; ,
\ee
where $g$ counts the number of internal degrees of freedom of the mother particle.
According to eq.~(\ref{relic}) this corresponds to $ Y_{Z_{\mu\tau}}= 0.3 \times 10^{-10} $
and is in perfect agreement with the plateau in Fig.~\ref{3a}. 
One interesting point to note is that as we increase the $\zmt$ mass $M_{\zmt}$, 
keeping $ g_{\mu\tau} $ fixed, the DM abundance decreases instead of increasing, as 
explained by the relation above. 
In the same figure also the production of dark matter as a result of the out-of-equilibrium decay 
of $\zmt$ can be seen beautifully. 
Less production of $\zmt$ results in lower DM abundance, since practically every $\zmt$ produces
two Dark Matter particles.

\begin{figure}[h!]
\centering
\includegraphics[angle=0,height=7cm,width=8cm]{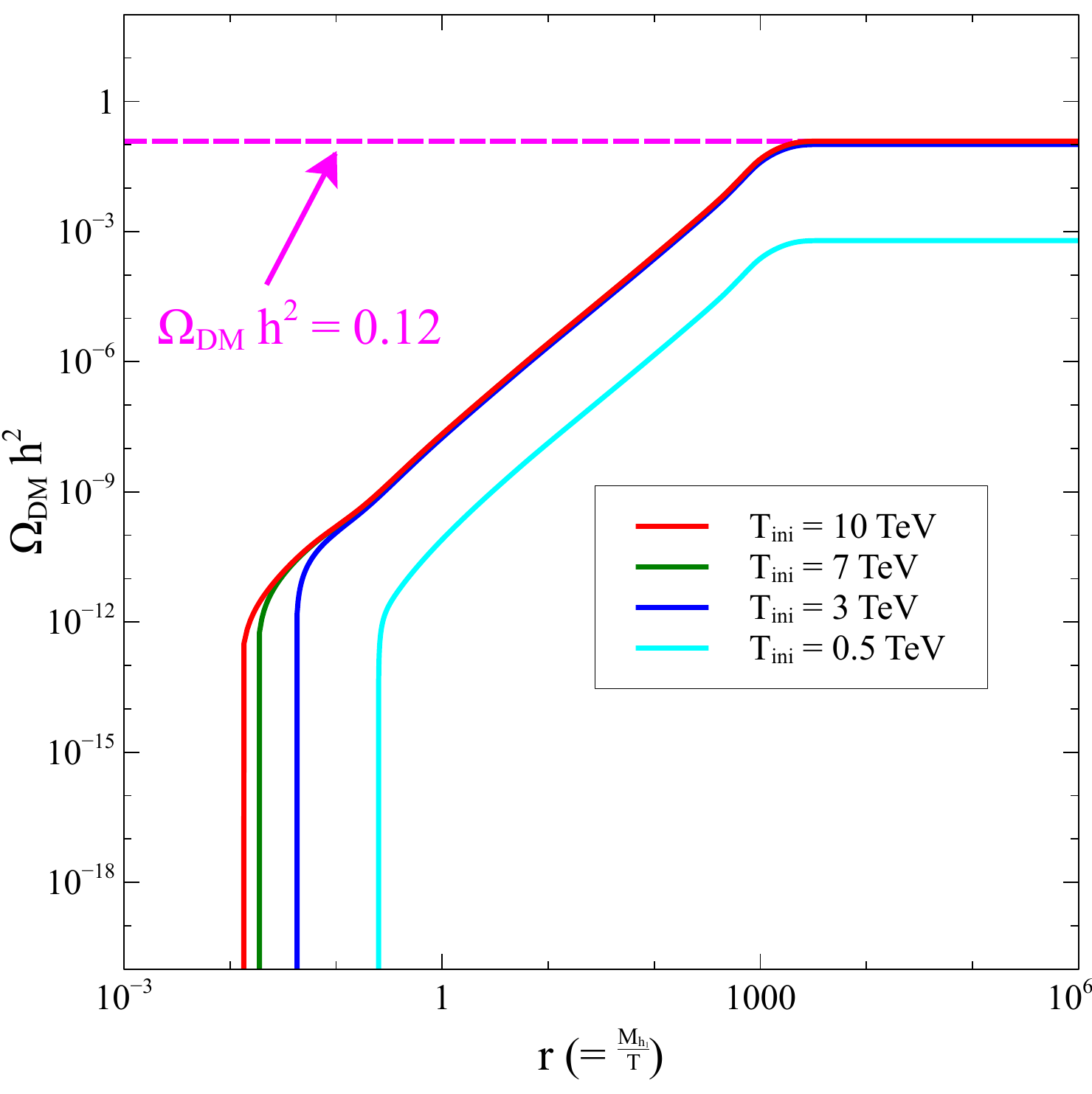}
\includegraphics[angle=0,height=7cm,width=8cm]{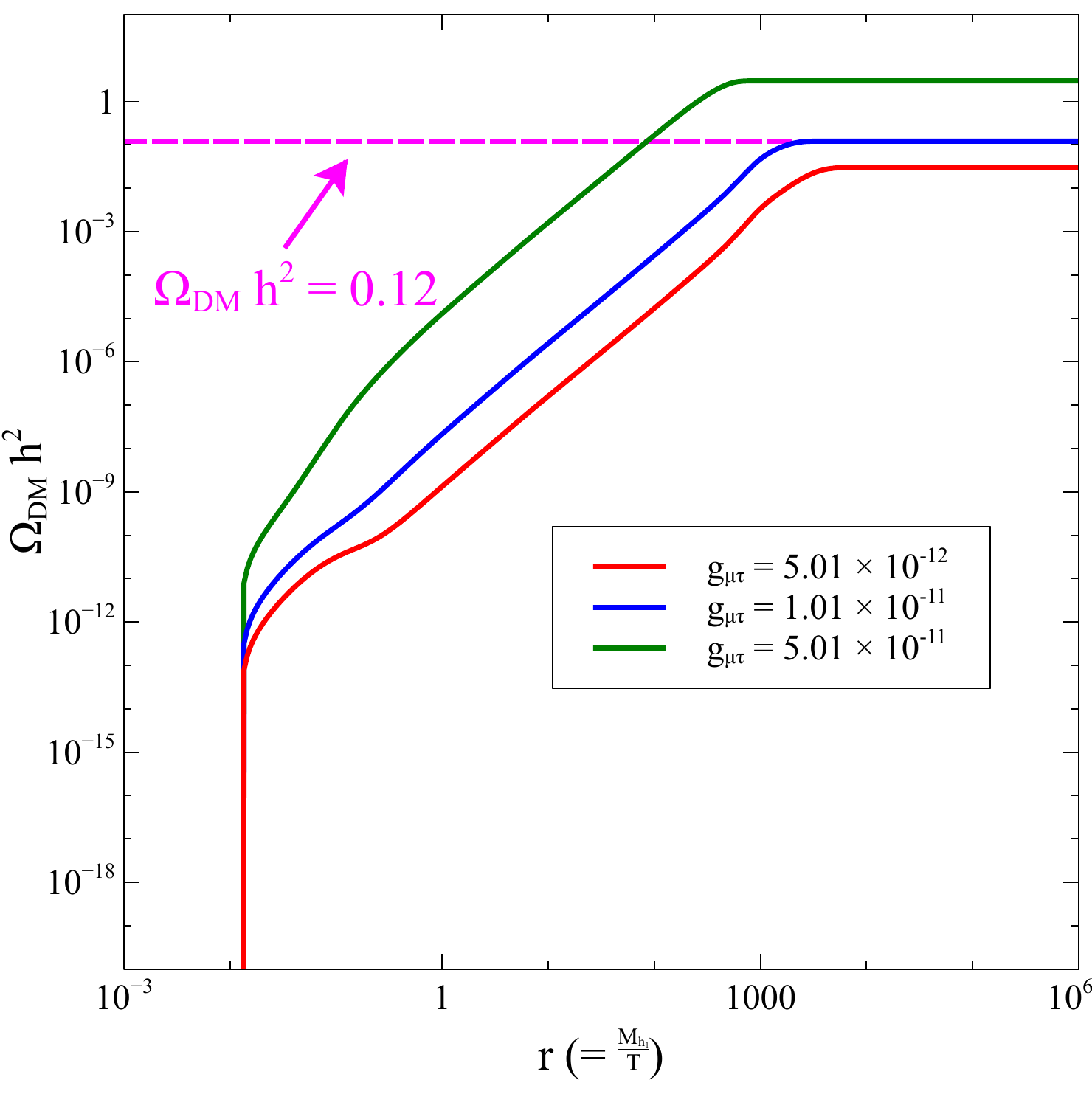}
\caption{Left (Right) panel: Variation of relic density with $r$ for different initial
temperature (for different gauge coupling values), while the other parameters
have been kept fixed at $g_{\mu\tau} = 1.01 \times 10^{-11}$ ($T_{ini} = 10$ TeV),
mixing angle $\alpha = 0.01$, gauge boson mass $M_{\zmt} = 1$ TeV, BSM Higgs mass
$M_{h_2} = 5$ TeV and RH neutrinos masses $M_{N_1}$ = 150 GeV, $M_{N_2} \simeq M_{N_3}$ =
100 GeV.}
\label{1a}
\end{figure}

The left panel of Fig.~\ref{1a} shows the variation of relic density with the parameter $r$
for different initial temperature $T_{ini}$ (temperature where DM relic density is taken as zero).
Important point to note here that as long as the initial temperature 
is above the mass of the gauge boson $Z_{\mu\tau}$, final relic density remains the same. However, 
when we reduce the initial temperature below the $\zmt$ mass (shown by the cyan color curve) then final abundance reduces significantly
due to the Boltzmann suppression factor. In the right panel 
we show the variation of DM relic density with $r$ for different gauge coupling
values ($g_{\mu\tau}$). One can see from the figure that if we increase
the value of the gauge coupling, the DM production rate as well as the total DM abundance 
increases. The reason can be understood from Eq.~(\ref{dkz}) which shows that the DM production rate, which is almost the same as the 
$Z_{\mu\tau}$ decay rate, is proportional to the second power of $\gmt$. 
In the present model for $\gmt = 1.01 \times 10^{-11}$
we achieve the correct DM relic density value of the universe. 
In both the panels of Fig.~\ref{1a}, the horizontal
magenta line corresponds to the present day correct DM relic density value of the universe.
For the rest of the analysis, we have fixed the initial temperature of the universe at 10 TeV.   
  

\begin{figure}[h!]
\centering
\includegraphics[angle=0,height=7cm,width=8cm]{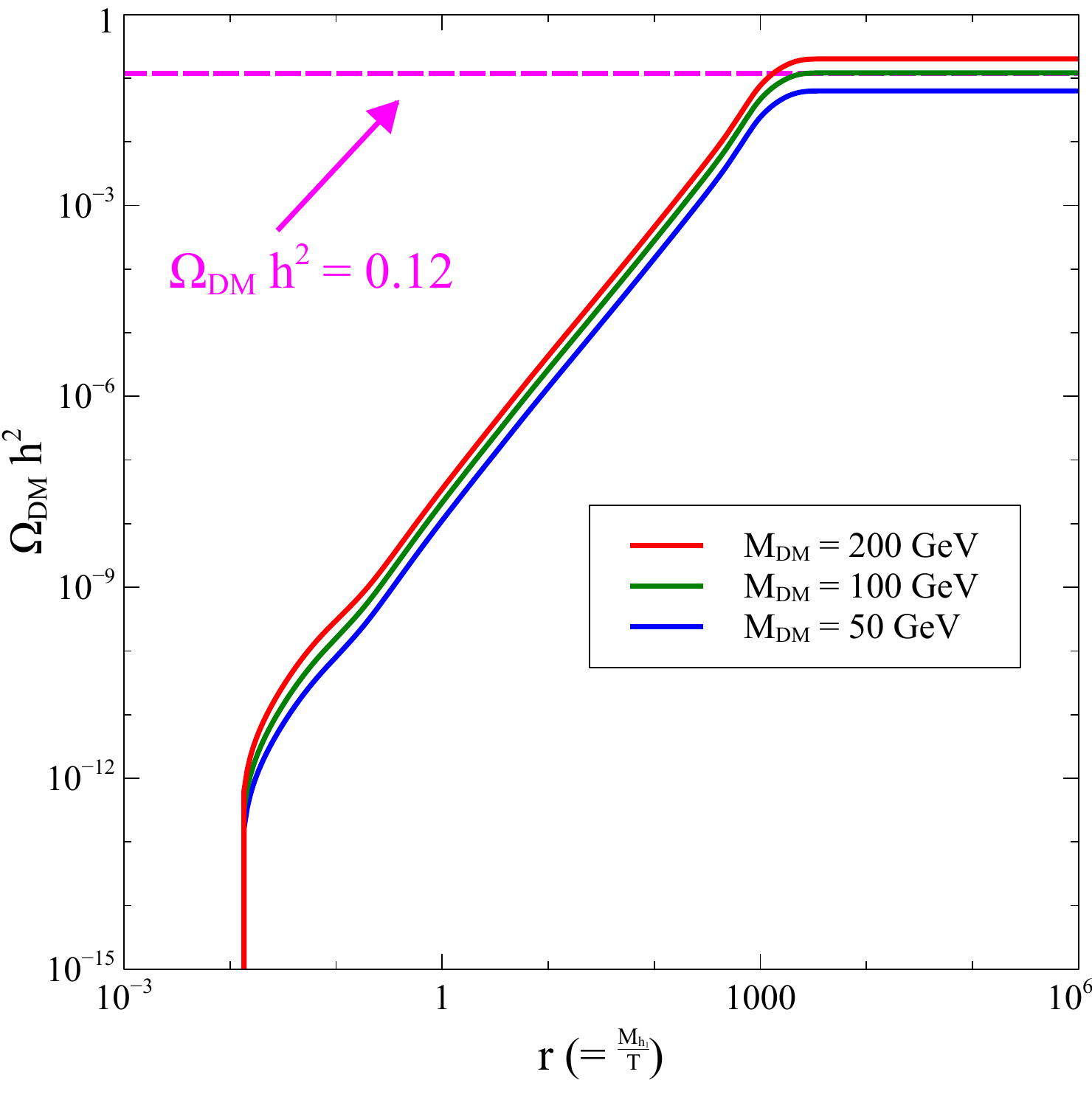}
\includegraphics[angle=0,height=7cm,width=8cm]{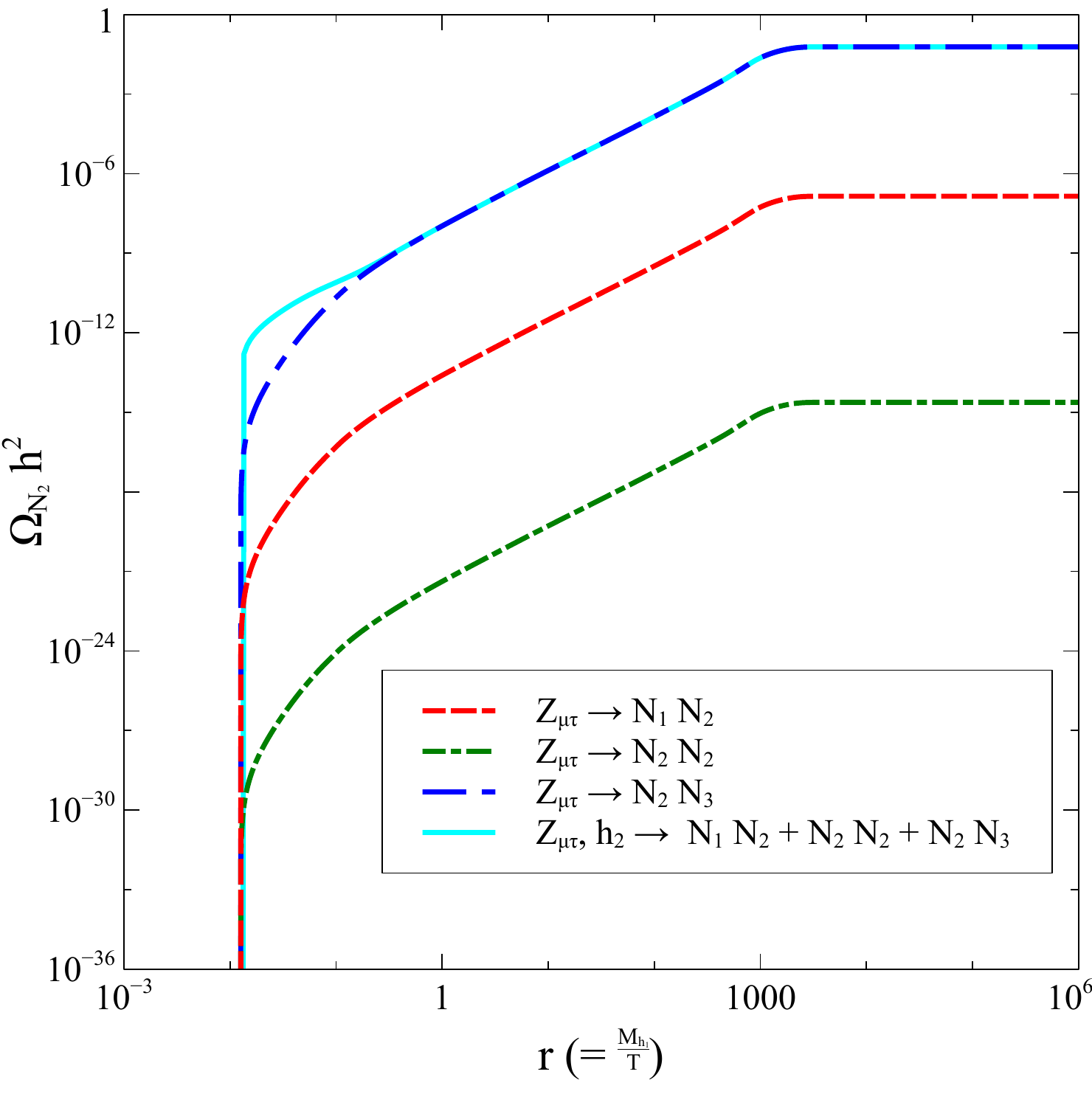}
\caption{Left (Right) panel: Variation of relic density with $r$ for different
values of DM mass (Contributions
in the relic density of DM from different channels of $\zmt$), while the other parameters
have been kept fixed at $g_{\mu\tau} = 1.01 \times 10^{-11}$,
mixing angle $\alpha = 0.01$, gauge boson mass $M_{\zmt} = 1$ TeV ($M_{DM} = 100$ GeV),
BSM Higgs mass $M_{h_2} = 5$ TeV and RH neutrinos masses
$M_{N_1}$ = 150 GeV, $M_{DM}=M_{N_2} \simeq M_{N_3}$ = 100 GeV.}
\label{2a}
\end{figure}

In the left panel of Fig.~\ref{2a}, we present the variation of the DM relic density for
three different values of the DM mass $M_{DM}$ (=$M_{N_2}, M_{N_3}$). As shown in
Eq.~(\ref{relic}) that DM relic density is proportional to the DM mass $M_{N_2}$
and $M_{N_3}$ and this dependence is evident in the left panel of Fig.~\ref{2a}. 
For the chosen value of
the parameters (mentioned in the caption), we have obtained correct relic density
value (indicated by the horizontal line) of the universe for DM mass value
$M_{DM} = M_{N_2} \simeq M_{N_3}$ = 100 GeV, this value will be different for
different set of values of the other parameters. 
In the right panel of Fig.~\ref{2a}, we show the decay contributions of $\zmt$
in different channels. The relative contributions among the different channels 
is seen to differ significantly and the decay rate into
$N_2 N_3$ dominates naturally producing equal populations of the two Dark
Matter candidates. 
Indeed, to produce degenerate neutrinos i.e. $M_{N_2} \simeq M_{N_3}$, we have
considered relatively small values of $\frac{h_{e\mu} v_{\mu\tau}}{\sqrt{2}}$
and $\frac{h_{e\tau} v_{\mu\tau}}{\sqrt{2}}$ ($\sim 0.1$), as discussed before. 
Therefore, the elements of the unitary matrix which relate the flavour and mass 
basis of the RH neutrinos take the following form,
$U_{11} \sim 1$, $U_{12}, U_{13}, U_{21}, U_{31} \sim 0.01$,
$U_{22} = U_{23} = \frac{1}{\sqrt{2}}$ and $U_{32} = - U_{33} = -\frac{1}{\sqrt{2}}$.
Therefore, it is clear from the couplings (as listed in Eq.~(\ref{zmt_coup})) that
the dominant channel for DM production is $Z_{\mu\tau} \rightarrow
N_{2} N_{3}$, while the other channels will be suppressed which is clearly 
visible in the right panel of Fig.\,\ref{3a}. Similar considerations will 
also be true for the $N_3$ DM production channels.

\section{3.5 keV $\gamma$ ray line}
\label{355-part}
\begin{figure}[h!]
\centering
\includegraphics[angle=0,height=5.5cm,width=14.0cm]{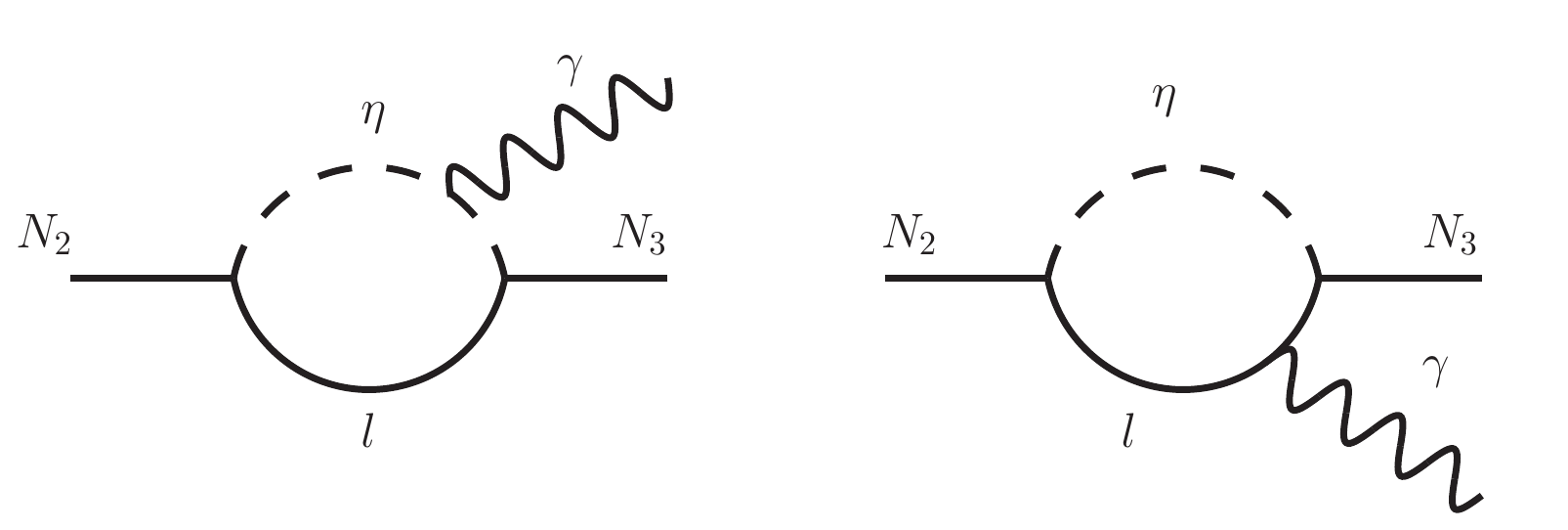}
\caption{Radiative decay of RH neutrino ($N_{2} \rightarrow N_{3} \, \gamma$) and 3.55 keV $\gamma$-line.}      
\label{fig_3.55}
\end{figure}
Finally, we come to the explanation of the $3.5$ keV $\gamma$-ray line from the RH neutrino 
radiative decay $N_{2} \rightarrow N_{3} \, \gamma$. 
Since the photon flux for a decaying Dark Matter candidate is given by
\be
\Phi = \frac{1}{4\pi M_{N_2} \tau_{N_2}} \int_{l.o.s.} \rho_{N_2}(\vec{r}) d\vec{r}
\ee
where the last integral over the $N_2$ density is computed along the line of sight
and $\tau_{N_2}$ is the lifetime of the heavier DM particle $N_2$.
In order to explain the 3.5 keV line from a decay such as $N_{2} \rightarrow N_{3} \, \gamma$, 
we need not only a mass splitting between the two fermion states of $ \sim 3.5$ keV, but
also a decay width of the unstable DM given as,
\begin{eqnarray}
\Gamma(N_2 \rightarrow N_{3} \gamma) = (0.72-6.6) \times
10^{-52}\, {\rm GeV}\, \left(\frac{M_{N_2}}{3.5\, {\rm keV}}\right)
= (0.2-1.9) \times 10^{-44}\, {\rm GeV}\, \left(\frac{M_{N_2}}{100\, {\rm GeV}}\right)
\,.
\end{eqnarray}
Here we are assuming that the density of $N_2 $ is approximately half of the DM density
and rescaled the result of \cite{Bulbul:2014sua} accordingly.

The relevant decay diagrams for $ N_2$ are shown in Fig.~\ref{fig_3.55}. 
We consider $N_2$ to be slightly heavier than $N_{3}$ ($\sim 3.5$ keV) so that it 
can produce the $3.5$ keV $\gamma$-ray line. As discussed before, the 3.5 keV mass 
splitting between nearly-degenerate $N_2$ and $N_3$ can be easily achieved in our model 
via the $\umt$ symmetry and its breaking parameters.  So we take
$V_{e\alpha} = \frac{h_{e\alpha} v_{\mu\tau}}{\sqrt{2}} \sim 0.1 $ GeV ($\alpha = \mu, \tau$)
and by suitably adjusting the $V_{e\alpha}$ parameters we can generate the 3.5 keV mass 
gap between $N_2$ and $N_3$.
For the $\umt$ conserving leading terms in Eq.~(\ref{mncomplex}) we take the values 
$M_{ee}$ = 11 TeV and $M_{\mu\tau}$ = 100 GeV 
which gives us $M_{N_2}$ and $M_{N_3}$ $\sim 100$ GeV with 
opposite CP parities \cite{Pal:1981rm}. 
Ref. \cite{Pal:1981rm} has pointed out that if $N_{2}$
and $N_{3}$ have opposite CP, then the transition
from $N_{2}$ to $N_{3}$ is governed only by the magnetic moment term ($\mu_{23}$),
generated at one loop level as shown in Figure~\ref{fig_3.55}. 
Therefore, the effective Lagrangian for the decay process 
$N_{2} \rightarrow N_{3} \, \gamma$ is given as 
\begin{eqnarray}
\mathcal{L}_{eff} \approx i\, \frac{\mu_{23}}{2}\,
\bar{N_{2}}\, \sigma^{\mu \nu} N_{3}\, F_{\mu \nu}
\,.
\end{eqnarray}
In determining the expression for the above decay process we consider the 
ratio of lepton mass to RH neutrino mass to be very small ($\frac{M_{l}}{M_{N_{2}}}\ll 1$). 
Also, the ratio of the RH neutrino mass and the inert doublet mass
is very small i.e. $\frac{M_{N_{2}}}{M_{\eta}}\ll 1$. The decay width of $N_2$ comes out as 
\cite{Garny:2010eg},  
\begin{eqnarray}
\Gamma(N_{2} \rightarrow N_{3} \gamma) = \frac{\mu_{23}^{2}}{4 \pi}\,
\delta^{3}\, \left(1 - P\,\frac{M_{N_3}}{M_{N_2}}\right)^{2}
\,,
\label{dm-decay}
\end{eqnarray} 
where $\delta = \frac{M_{N_2}}{2}
(1 - \frac{M_{N_3}^{2}}{M_{N_2}^{2}})$, $P$ gives the 
relative CP of the two neutrino states, which 
in the present model is $P= -1$. The magnetic moment coefficient $\mu_{23}$ 
in our model is given by
\begin{eqnarray}
\mu_{23} = \sum_{i} \frac{e}{2} \frac{1}{(4 \pi)^{2}} \frac{M_{N_2}}{M_{\eta}^{2}}
(y_{i 2} y_{i 3})
\,,
\label{mag}
\end{eqnarray} 
where $y_{ij}=h_iU_{ij}$ being the derived Yukawa couplings given in Eq.~(\ref{yuk}) . 
The values of the parameters appearing in the $N_2$ decay width are intimately related with 
those that determine the light neutrino masses. 
In Section \ref{neutrino-part}, we had set the parameter values to explain the tiny
neutrino mass in the following order,
\begin{eqnarray}
M_{\eta} = 10^{6}\, {\rm GeV}, M_{N_2} = 100\, {\rm GeV}, (y_{ij})^{2} = 10^{-1} 
\,.
\end{eqnarray}
Using these in the Eq.~(\ref{yuk}) we get $\mu_{23} \sim \mathcal{O}(10^{-14})$ GeV$^{-1}$. 
Using Eq.~(\ref{dm-decay}), for DM mass around 100 GeV, $\delta \simeq 3.5$ keV and 
$\mu_{23} \sim 10^{-14}$ GeV$^{-1}$, we get the lifetime of $N_2$ of the order 
$\mathcal{O}(10^{-44})$ GeV, which is exactly what is needed to give the 3.5 keV line.
Note that the lifetime of $N_2$ is then around $10^{19}$ sec and hence greater 
than the age of the universe ($10^{17}$ sec).
Hence the present model can naturally explain the origin of the claimed 3.5 keV line.

\section{Conclusion}
\label{con}

In the present work we extended the SM gauge group by a local $U(1)_{L_{\mu} - L_{\tau}}$ 
gauge group and a $\mathbb{Z}_{2}$ discrete symmetry.
The particles spectrum was extended by three RH neutrinos, one inert doublet and
one SM gauge singlet scalar. 
We showed that this model explains the observed 3.5 keV line consistently with the relic dark matter abundance in the framework 
of a model that generates light neutrino masses radiatively. The Type I seesaw in this model is forbidden by the $\mathbb{Z}_{2}$ symmetry but  
tiny neutrino masses are generated via a one-loop diagram involving the RH neutrino and the inert doublet which does not take any VEV. We considered inert scalar masses $\sim 10^6$ GeV, which is higher than the reheat temperature, and RH neutrino masses $\sim 100$ GeV. Then for parameter choices $\lambda_5\sim 10^{-3}$ and Yukawa couplings $y_{ji}^2 \sim 10^{-1}$ we can get light neutrino masses $M_\nu \sim 0.01$ eV. The RH neutrino mass matrix in our model is non-diagonal and carries the $L_{\mu} - L_{\tau}$ flavour structure which ensures that two of the RH neutrino remain degenerate in the $U(1)_{L_{\mu} - L_{\tau}}$ symmetric limit. The spontaneous breaking of the $U(1)_{L_{\mu} - L_{\tau}}$ gauge symmetry generates terms in the RH neutrino mass matrix that splits the two degenerate RH neutrinos by 3.5 keV, while the third one remains heavier. The two nearly degenerate neutrinos form the two-component DM in our model. We showed that the RH neutrinos are predominately produced by the decay of the extra neutral gauge boson $Z_{\mu\tau}$, which are taken in the 1 TeV mass range in our model. The production of RH neutrinos from decay of the additional scalar $h_2$ is subdominant, while the annihilation channels have negligible effect. We showed that the peculiar structure of the unitary matrix ($U$) which relates the flavour and mass basis of the RH neutrinos ensures that the decay mode $\zmt \rightarrow N_{2}\,N_{3}$ is the dominant one among the other channels. Since the associated gauge coupling $g_{\mu\tau}$ is taken to be very small here, the $Z_{\mu\tau}$ stays out of equilibrium in the early universe and the RH neutrinos are produced by the freeze-in mechanism. We solved the coupled Boltzmann equation numerically and showed the dependence of the DM relic abundance on initial temperature $T_{ini}$, $g_{\mu\tau}$, $M_{\zmt}$ and $M_{DM}$. Finally, we showed that the heavier of the two DM component $N_2$ can decay into the lighter $N_3$ ($N_{2} \rightarrow N_{3}\,\gamma$) through one loop diagram, thus producing the 3.5 keV X-ray line that was observed by Chandra satellite. The model parameter values which determine the lifetime of $N_2$ were obtained through constraints from the light neutrino mass sector 
and gave a decay rate of $10^{-44}$ GeV for $N_2$. So the lifetime of the heavier Dark Matter particle is consistent 
with both the age of the universe as well as the strength of the observed 3.5 keV line. 

Regarding collider observables, this model unfortunately does not give many promising signatures.
Indeed all the particles of the gauged $ \mu-\tau $ sector interact with the Standard Model only via the very
small coupling $\gmt \sim 10^{-11} $, so that their production at LHC or their effect on precision
observables is very suppressed.  If one would be able to produce those states, a long lifetime and possibly
displaced vertices could be the characteristic signature~\cite{Falkowski:2014sma, Arcadi:2014tsa, Arcadi:2014dca}.
On the other hand, more substantial can be the production cross-section for the heavier Higgs boson $ h_2$, 
depending on its mass the mixing angle $\alpha $. Unfortunately in this case, its dominant decay channels 
are those in Standard Model states through the mixing with the Higgs doublet and so the connection of this 
heavy state with the neutrino sector and the $ U(1)_{L_\mu-L_\tau} $ will be difficult to prove.

\section{Acknowledgements} 
The authors would like to thank the Department of Atomic Energy
(DAE) Neutrino Project under the XII plan of Harish-Chandra
Research Institute. 
LC would like to thank the Harish-Chandra Research Institute
for hospitality during the initial stages of this work.
This project has received funding from the European Union's Horizon
2020 research and innovation programme InvisiblesPlus RISE
under the Marie Sklodowska-Curie
grant  agreement  No  690575. This  project  has
received  funding  from  the  European
Union's Horizon  2020  research  and  innovation
programme  Elusives  ITN  under  the 
Marie  Sklodowska-
Curie grant agreement No 674896.

\appendix
\section*{Appendix}
\section{Analytical Expression of relevant Decay width and Collision terms}
\label{App:AppendixA}
If we consider a generic process $\chi (\tilde{p}) \rightarrow a(\tilde{p_1})\,
b(\tilde{p_2})$ (where $\tilde{p} = (E_{p},\bar{p})$)
then the collision term will take the
following form \cite{Kolb:1990vq, Gondolo:1990dk},
\begin{eqnarray}
\mathcal{C}[f_{\chi}(p)]&=&\dfrac{1}{2\,E_p}\int
\dfrac{g_a\,d^{3} p1}{(2\pi)^{3} \,2E_{p1}}
\dfrac{g_b\,d^{3} p2}{(2\pi)^{3} \,2E_{p2}}
(2\pi)^4\,\delta^4(\tilde{p}-\tilde{p_1}-\tilde{p_2})
\times\overline{\,\lvert \mathcal{M}\rvert^2}\nonumber \\
&&~~~~~~~~~~~~~~\times\,[f_a\,f_b\,\left(1 \pm f_\chi\right)
-f_\chi\left(1 \pm f_a\right)\left(1 \pm f_b\right)]\,.
\label{colision1}
\end{eqnarray}
Now the full expressions of the collision terms in
Eq.~(\ref{zmt_prod}) are as
follows \cite{Konig:2016dzg, Biswas:2016iyh},
\begin{itemize}
\item $\mathcal{C}^{\zmt \rightarrow all}$:
\,\,Collision term for the extra gauge boson $\zmt$ decay can be written in
the following way in terms of the parameters which we have introduced
in Section \ref{bee}. 
\begin{eqnarray}
\mathcal{C}^{\zmt \rightarrow all}&=&-f_{\zmt}(\xi_p)\times
\Gamma_{\zmt \rightarrow all}\times
\dfrac{r_{\zmt}}{\sqrt{\xi_p^2\,\mathcal{B}(r)^2+r_{\zmt}^2}}\,.
\end{eqnarray}
where $\Gamma_{\zmt \rightarrow all} = \Gamma_{\zmt \rightarrow f \bar{f}}$
+ $\Gamma_{\zmt \rightarrow N_{i} N_{j}}$ and the expression for the each decay terms
are as follows,
\begin{eqnarray}
\Gamma_{\zmt \rightarrow f \bar{f}} &=&
\frac{M_{\zmt} \,g_{\mu\tau}^{2}}{12\,\pi}
\left( 1 + \frac{2 M_{f}^{2}}{M_{\zbl}^{2}} \right)
\sqrt{1 - \frac{4 M_{f}^{2}}{M_{\zbl}^{2}}} \nn \\ 
\Gamma_{\zmt \rightarrow N_{i} N_{j}} &=& \frac{M_{\zmt}
g_{\zmt N_{i} N_{j}}^{2}}{12\,\pi S_{ij}}
\left(1 - \frac{(M_{N_i}+M_{N_j})^{2}}{M_{\zmt}^{2}} \right)^{3/2} \nn \\
&& \times \left(1 - \frac{(M_{N_i} - M_{N_j})^{2}}{M_{\zmt}^{2}} \right)^{1/2}
\times \left(1 - \frac{(M_{N_i} - M_{N_j})^{2}}{2\,M_{\zmt}^{2}} \right)
\label{dkz}
\end{eqnarray} 
where $f\, =\,\nu_{\mu},\nu_{\tau},\mu^{\pm}$ and $\tau^{\pm}$
because of the $(L_{\mu} - L_{\tau})$ symmetry of the present model
and the couplings take the following form depending on RH neutrinos,
\begin{eqnarray}
g_{\zmt N_{2} N_{2}} &=& -\frac{\gmt}{2}\,(U_{22}^{2} - U_{32}^{2}) \nn \\
g_{\zmt N_{3} N_{3}} &=& -\frac{\gmt}{2}\,(U_{23}^{2} - U_{33}^{2}) \nn \\
g_{\zmt N_{1} N_{2}} &=& -\frac{\gmt}{2}\,(U_{21} U_{22} - U_{31} U_{32}) \nn \\
g_{\zmt N_{1} N_{3}} &=& -\frac{\gmt}{2}\,(U_{21} U_{23} - U_{31} U_{33}) \nn \\
g_{\zmt N_{2} N_{3}} &=& -\frac{\gmt}{2}\,(U_{22} U_{23} - U_{32} U_{33})
\label{zmt_coup}
\end{eqnarray}

The statistical factor $S_{ij} = 2$ for i = j and 1 for i $\neq$ j.
\item $\mathcal{C}^{h_{2} \rightarrow \zmt \zmt}$:\,\,
The collision term for the extra geuge boson production $\zmt$
from the decay of BSM Higgs $h_2$ takes the following form,
\begin{eqnarray}
\mathcal{C}^{h_2 \rightarrow \zmt\zmt} &=&
\dfrac{r}{8\pi M_{sc}}\dfrac{\mathcal{B}^{-1}(r)}
{\xi_p \sqrt{\xi_p^2\mathcal{B}(r)^2+
\left(\dfrac{M_{\zbl}\,r}{M_{sc}}\right)^2}}
\dfrac{g_{h_2\zmt\zmt}^2}{6}
\left(2+\dfrac{(M_{h_2}^2-2M_{\zmt}^2)^2}{4M_{\zmt}^4}\right)\nonumber \\ 
&&\times \left(e^{-\sqrt{\left(\xi_{k}^{\rm min}\right)^2
\mathcal{B}(r)^2+\left(\frac{M_{h_2}\,r}{M_{sc}}\right)^2}}
\,-\,e^{-\sqrt{\left(\xi_{k}^{\rm max}\right)^2
\mathcal{B}(r)^2+\left(\frac{M_{h_2}\,r}{M_{sc}}\right)^2}}
\right) \,.
\label{ch2zblzbl-final}
\end{eqnarray}

where 

\begin{eqnarray}
g_{h_{2}{\zmt} \zmt} &=& \frac{2 M_{\zmt}^{2} \cos \alpha}{v_{\mu \tau}}, \nn \\
\xi_k^{\rm min} (\xi_p,r)&=&\dfrac{M_{sc}}{2\,\mathcal{B}(r)\,r\,M_{\zmt}}
\left| \,\eta (\xi_p,r)-\dfrac{\mathcal{B}(r)
\times M_{h_2}^2}{M_{\zmt} \times M_{sc}}\,\xi_p\,r
\right| \,,\nn \\
\xi_k^{\rm max} (\xi_p,r)&=&\dfrac{M_{sc}}{2\,\mathcal{B}(r)\,r\,M_{\zmt}}
\bigg( \,\eta (\xi_p,r)+\dfrac{\mathcal{B}(r)
\times M_{h_2}^2}{M_{\zmt}\times M_{sc}}
\,\xi_p\,r \,\bigg)\,,\nn \\
\eta(\xi_p,r)&=& \left(\frac{M_{h_2}\,r}{M_{sc}}\right)
\,\sqrt{\dfrac{M_{h_2}^2}{M_{\zmt}^2}-4}\,\,
\sqrt{\xi_p^2\,\mathcal{B}(r)^2+
\left(\frac{M_{\zmt}\,r}{M_{sc}}\right)^2}\,.
\end{eqnarray}

\item $\Gamma_{h_{k} \rightarrow N_{i} N_{j}}$: \,\,Decay width for 
the SM like Higgs ($h_1$) and BSM Higgs ($h_2$) take the following form,
\begin{eqnarray}
\Gamma_{h_k \rightarrow N_{i} N_{j}} &=&
\frac{M_{h_k} g_{h_{k} N_{i} N_{j}}^{2}}{8 \,\pi S_{ij}}
\left( 1 - \frac{(M_{N_i} + M_{N_j})^{2}}{M_{h_k}^{2}} \right)^{3/2} \nn \\
&& \times \left( 1 - \frac{(M_{N_i} - M_{N_j})^{2}}{M_{h_k}^{2}} \right)^{1/2} 
\label{dkh}
\end{eqnarray} 
where the couplings take the following form,
\begin{eqnarray}
g_{h_{2\,(1)} N_{1} N_{2}} &=& - \frac{\sqrt{2} \cos \alpha\, (\sin \alpha)}
{4}\, (U_{11} U_{22}
h_{e\mu} + U_{12} U_{21} h_{e\mu} + U_{11} U_{32} h_{e\tau} +
U_{12} U_{31} h_{e\tau})\,\,\,\nn \\
g_{h_{2\,(1)} N_{1} N_{3}} &=& - \frac{\sqrt{2} \cos \alpha\, (\sin \alpha)}
{4}\, (U_{11} U_{23}h_{e\mu} + 
U_{13} U_{21} h_{e\mu} + U_{11} U_{33} h_{e\tau} +
U_{13} U_{31} h_{e\tau})\,\,\,\nn \\
g_{h_{2\,(1)} N_{2} N_{3}} &=& - \frac{\sqrt{2} \cos \alpha\, (\sin \alpha)}
{4}\, (U_{12} U_{23}
h_{e\mu} + U_{13} U_{22} h_{e\mu} + U_{12} U_{33} h_{e\tau} +
U_{13} U_{32} h_{e\tau})\,\,\,\nn \\
g_{h_{2\,(1)} N_{2} N_{2}} &=& - \frac{\sqrt{2} \cos \alpha\, (\sin \alpha)}
{2}\, (U_{12} U_{22} h_{e\mu} + U_{12} U_{32} h_{e\tau})\,\,\,\nn \\
g_{h_{2\,(1)} N_{3} N_{3}} &=& - \frac{\sqrt{2} \cos \alpha\, (\sin \alpha)}
{2}\, (U_{13} U_{23} h_{e\mu} + U_{13} U_{33} h_{e\tau})\,\,\,
\end{eqnarray}

\end{itemize}


\end{document}